# III. Geometrical framework for thinking about globular proteins: turns in proteins

Short running title (max. 40 characters): Turns in globular proteins


Tatjana Škrbić[1,2], Achille Giacometti[1,3], Trinh X. Hoang[4], Amos Maritan[5], and Jayanth R. Banavar[2]

[1] Ca' Foscari University of Venice, Department of Molecular Sciences and Nanosystems, Venice, Italy

[2] University of Oregon, Department of Physics and Institute for Fundamental Science, Eugene, Oregon, USA

[3] European Centre for Living Technology (ECLT), Ca' Bottacin, Dorsoduro 3911, Calle Crosera, 30123 Venice, Italy

[4] Vietnam Academy of Science and Technology, Institute of Physics, Hanoi, Vietnam

[5] University of Padua, Department of Physics and Astronomy, Padua, Italy

**Correspondence**

Tatjana Škrbić
Ca' Foscari University of Venice, Department of Molecular Sciences and Nanosystems, Via Torino 155, 30170 Venice, Italy.
**Email: tatjana.skrbic@unive.it**





**Abstract**

We have shown recently that the notion of poking pairwise interactions along a chain provides a unifying framework for understanding the formation of both secondary and the tertiary protein structure based on symmetry and geometry. α-helices and β-sheets are found to be special geometries that have systematic poking contacts in a repetitive manner with the contacts being local along the α-helix and non-local along a pair of adjacent strands within a β-sheet. Pairwise poking interactions also govern tertiary structure formation, but they are weaker and there are no special geometrical constraints as in secondary structure formation. Here we demonstrate that protein turns, the most prevalent non-repetitive structural element in proteins, are instances of local (as in α-helices) and isolated (non-repetitive) poking pairwise contacts for which the geometrical constraints are partially relaxed. This simple and purely geometrical definition of protein turns (also sometimes known as reverse turns, β-turns, β-bends, hairpin bends, $3_{10}$ bends, kinks, widgets, ...) provides a simple framework for unifying them. We present the results of a


systematic analysis and identify their structural classes as well as their respective amino acid preferences.

*Keywords:* geometry, symmetry, poking, backbone, amino acids



## Statement for broader audience

Poking interaction along a discrete chain signals the affinity of one part of the chain with another. It has been shown that a simple and purely geometrical model based on poking interactions is able to capture the essential features of complex protein chains, especially their building blocks and their assembly. We demonstrate that poking interactions also provide a simple framework to describe protein turns as isolated (not coordinated) local poking interactions. This framework allows one to comprehend the structural variety of protein turns that have been extensively studied in the past.



## 1. Introduction

The building blocks of the structure of globular proteins, topologically one-dimensional α-helices [1] and almost planar β-sheets [2] are lower dimensional manifolds and thus not subject to compactification in three-dimensional space. Furthermore, both a helix and a strand are iso-directional. For the formation of a reasonably compact tertiary structure of a globular protein, one requires a sharp turn allowing for a change in the chain direction [3-28]. Indeed, after helices and sheets, protein turns are the most common structural feature recognized in proteins [8,14]. In contrast to helices and sheets, they are non-repetitive structural elements with highly heterogeneous structures, that have made them, despite their ubiquity, much more challenging to define and categorize [27]. Protein turns are prevalently located at the surface of a protein and facilitate interactions with other molecules and often serve as sites for high specificity ligand binding [20,27]. Having a high preference for hydrophilic residues [13,15], because of their location at the protein surface and being exposed to water, turns help with the overall stabilization of the three-dimensional native state of a globular protein [21,25]. Protein turns are thus special in their ability to provide for the combined structural and chemi-



cal backdrop [26]: tight turns bring non-local parts of a protein chain together and at the same time have highly specific amino acid composition [8,13-15,20,27]. In this way, they allow for an efficient navigation of complex folding pathways and play an important role in protein folding and function [4,10,19,22,28].

Protein turns go under various names as reverse turns, β-turns, β-bends, hairpin bends, $3_{10}$ bends, kinks, widgets, ... and they are found to display an enormous structural variety [8,9,14,16]. Consequently, there are many distinct definitions of protein turns present in the literature. Here we seek a physically motivated unified perspective for understanding turns.

The structural characterization and classification of protein turns was presented in the seminal work of Venkatachalam [3], who explored the conformations available to a system of three linked peptide units (i.e., to a consecutive amino acid quartet) that could be stabilized by a backbone hydrogen bond between the -CO carboxyl group of the i-th residue and the -NH amino group of the (i+3)-th residue. Ventakachalam [3] suggested three turn types, named Type I, Type II and Type III (G-helix or $3_{10}$-turn), as well as their mirror images (Types I', II' and III', respectively) and determined their respec-



tive backbone conformations, in terms of the Ramachandran angles [29] $(\varphi_{i+1}, \Psi_{i+1})$ and $(\varphi_{i+2}, \Psi_{i+2})$ of the two inner residues i+1 and i+2 of the protein quartet (i,i+1,i+2,i+3). Types I and II (as well as their mirror images I' and II') are the most common turn types and are known as β-turns or U-turns, because they provide for the sharpest protein turns, as in β-hairpins where the polypeptide chain folds back on itself in a U-shape. Types III and III' are much milder turns and correspond to the turn in a G-helix and are thus called G-turns or $3_{10}$-turns. Two mirror image turns are transformed from one to the other by inversion symmetry with respect to the (0,0) point in the Ramachandran plot [29].

After extensive surveys of protein structures, it became increasingly evident that hydrogen bonding between residues i and i+3 was not a necessary condition to have a good turn. Furthermore, many turns were found to have the distance between the $C_\alpha$ atoms of the i and i+3 residues of up to ~7-7.5 Å lying further away than the plausible hydrogen bonding distance [6]. Thus, in addition to the three turn types of Ventakachalam and their mirror images [3], five additional turn types were identified that seemed to account for all observed cases [7,11,18].



Apart from the turn classification relying on Ramachandran angles and detailed atomistic conformational analysis, many authors have successfully adopted a coarse-grained approach considering only the $C_\alpha$ atoms. Kuntz [5] considered turns as all non-helical segments that effectively change the direction of the protein chain by more than 90°, while Levitt and Greer [12] assigned turns as non-α, non-β segments for which the dihedral angle defined by four successive $C_\alpha$ atoms is between -90° and 90°. In addition, Rose and Seltzer [9] introduced a structurally unbiased, purely geometrical definition of protein turns, and identified them as local minima in the radius of curvature calculated from the i-2, i, and i+2 $C_\alpha$ atoms and in which both the 'inlet' and the 'outlet' of the presumed turn cannot both be fitted within a single pipe of diameter ~5.2Å. Interestingly this estimate of Rose and Selzer is close to our recent theoretical prediction for the thickness of a discrete protein backbone of 2Δ~5.26Å, as measured by the diameter of a Kepler coin or twice the radius of curvature of a Kepler helix [30,31]. The Selzer-Rose pipe criterion is effective in eliminating spurious local minima within a single helix.



Recently, relying entirely on symmetry and geometry, and the notion of poking pairwise interactions within a discrete chain, we presented a unifying framework to understand the formation of both secondary and tertiary protein structure [30,31]. A poking interaction signals the affinity of one part of a chain to another and is often associated with a backbone hydrogen bond. The protein α-helix geometry is the unique way in which a pair of Kepler coins of a discrete chain at the i-th and (i+3)-th positions locally poke towards each other in a repetitive manner and just touch each other [30,31]. The turn angle of a quartet τ is defined as the angle between the directions (i,i+1) and (i+2,i+3). A cautionary note is that these two lines are skew and do not inter-sect. The α-helix is iso-directional consisting of successive quartets comprising tight 'Kepler turns' that are necessarily non-planar to avert steric clashes and have local turn angles of around 130° (see Figure 1).



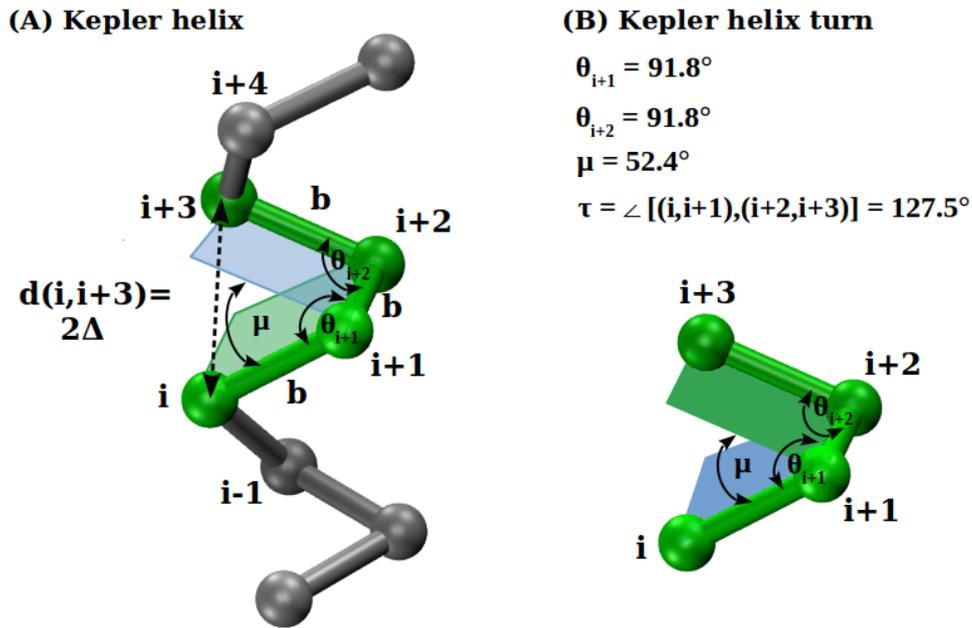

**(A) Kepler helix**

**(B) Kepler helix turn**

$\theta_{i+1} = 91.8°$
$\theta_{i+2} = 91.8°$
$\mu = 52.4°$
$\tau = \angle\ [(i,i+1),(i+2,i+3)] = 127.5°$

Figure 1: Panel (A) shows Kepler helix that consists of the consecutive local turns (Kepler helix turns), where one such turn is formed by quartet of points (i,i+1,i+2,i+3) and highlighted in green color. The bond length b=3.81Å sets the length scale, and the distance d(i,i+3) is equal to twice the value of the local radius of curvature of Kepler helix 2Δ. The geometrical characteristics of the Kepler turn, two bond bending angles $\theta_{i+1}$ and $\theta_{i+2}$ and dihedral angle $\mu$ are indicated. Panel (B) depicts enlarged Kepler turn, with the theoretically predicted values of its geometrical characteristics. Protein helices are characterized by a mean bond bending angle of 91.7±2.8°, a mean dihedral angle of



49.5±6.7°, and a mean local turn angle of 130.2±5.8°. These mean values and their standard deviations have been calculated over 226,478 α-helical quartets (for definition of an α-helical quartet see Materials and Methods, Section 2.4) in our data set (defined in Materials and Methods, Section 2.3).

Instead of strictly repetitive turns, as in the Kepler helix, let us consider a completely isolated poking contact between i and i+3 with the *absence* of a poking contact between i-1and i+2 *and* i+1 and i+4.  The absence of any constraint pertaining to the presence of a succession of pairwise poking contacts leads to a relaxion of the rigid constraint with the possibility of a close to planar configuration and thence a larger turn angle.  Our main goal is to demonstrate that protein turns are indeed instances of such local and isolated poking interactions.

## 2. Materials and Methods

## 2.1 Local protein geometry: protein quartets calculation



We view the protein backbone as a sequence of equidistant points, where the $C_\alpha$ atoms are located, separated by a constant bond length b of 3.81Å [32]. The shortest portion of protein local structure that can effectively 'turn' the chain consists of four consecutive points (quartet), see Figure 2. A truly perfect turn angle corresponds to $\tau = 180°$. A turn angle associated with just three amino acids is given by $\tau = 180°$- (the local bond bending angle). The latter angle is typically not much smaller than $90°$, so that a typical sharp triplet turn is around $90°$ (a right-angle turn) and is not effective in turning a chain back. This naturally leads to a quartet of beads, as in earlier studies, as the basic portion of protein local structure that serve as local turn candidates.

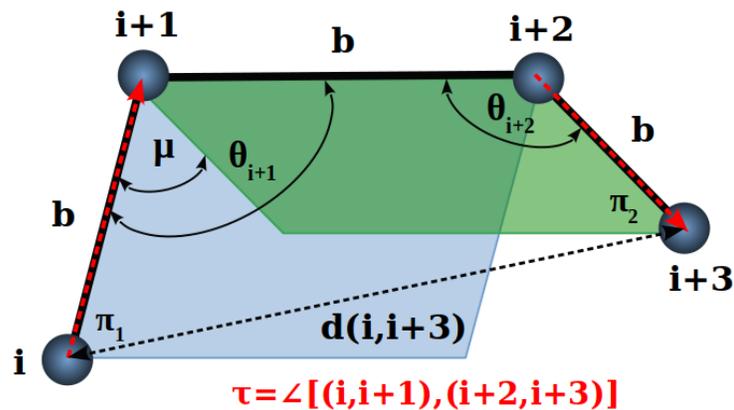

Figure 2: Definition of the local turn consisting of four successive points i, i+1, i+2, and i+3 along a chain where the $C_\alpha$ atoms are lo-



cated. The bond length b is the distance between the consecutive beads. The bond bending angle $\theta_{i+1}$ is the angle subtended at point (i+1) by points i and i+2, while the bond bending angle $\theta_{i+2}$ is the angle subtended at point i+2 by points i+1 and i+3. The dihedral angle μ is the angle between the planes $\pi_1$ and $\pi_2$ formed by [i,i+1,i+2] and [i+1,i+2,i+3], respectively. The distance between the first bead in the turn i and the last bead in the turn i+3, d(i,i+3), is indicated. The turn angle τ is defined as the angle between the directions (i,i+1) and (i+2,i+3). Thus, there are several variables that characterize a local quartet of amino acids with constant bond length b: $\theta_{i+1}$, $\theta_{i+2}$, μ, d(i,i+3), and τ, of which just three are independent.

Figure 2 depicts the quartet of beads and defines its relevant geometrical attributes. Under the conditions of constant bond length b, the geometry of the quartet is defined by exactly three independent variables. These three variables can be chosen from five convenient variables: the two bond bending angles $\theta_{i+1}$ and $\theta_{i+2}$, the dihedral angle μ, the distance d(i,i+3), and the turn angle τ (see Figure 2).



An important issue is the difference between unidirectionality and uni-axiality and its repercussions for the turn analysis we present in this paper. A protein chain is unidirectional at the atomic level because the C-terminal and N-terminal unambiguously denote the protein beginning and end. Uniaxiality, on the other hand, does not distinguish the direction in which the chain is going, as is the case when the geometries of the quartets [i,i+1,i+2,i+3] and [i+3,i+2,i+1,i] are exactly the same and the atomic structure is not considered. Because of this difference, we will be careful in all that follows to distinguish between the positions of the two inner quartet beads, (i+1) and (i+2), and consequently the bond bending angles $\theta_{i+1}$ and $\theta_{i+2}$, respecting the chain directionality.

We are after quartets of beads that form large turn angles $\tau$. Ideally, the maximum possible turn angle is $\tau_{max} = 180°$. To understand the general circumstances under which large turn angles are to be expected, we perform calculations in which we create all possible quartet geometries that satisfy steric constraints. In other words, we generate sequences of four points (i,i+1,i+2,i+3) with a constant bond length b=3.81Å, that yield a fixed distance d(i,i+3) between the ends of the quartet. In our calculations, for sim-



plicity, we place the second point of the quartet (i+1) at the origin of the coordinate system resulting in mathematical expressions for the quartet coordinates:

| | i=1 | i=2 | i=3 | i=4 |
|---|---|---|---|---|
| $x_i$ | $b \cos \theta_{i+1}$ | 0 | b | $b - b \cos \theta_{i+2}$ |
| $y_i$ | $b \sin \theta_{i+1}$ | 0 | 0 | $b \sin \theta_{i+2} \cos \mu$ |
| $z_i$ | 0 | 0 | 0 | $b \sin \theta_{i+2} \sin \mu$ |

The bond bending angles $\theta_{i+1}$ and $\theta_{i+2}$, are chosen in the restricted range [85°,180°], while the dihedral angle $\mu$ is unrestricted and can be in the range [-180°,180°] as in protein structures [32]. We find that there is a correlation between the degree of the planarity of the quartet making a turn (equivalent to the dihedral angle $\mu$ approaching 0°) and its turn angle $\tau$, as illustrated in Figure 3. For a given value of the dihedral angle $\mu$, there are distinct combinations of the two bond bending angles $(\theta_{i+1}, \theta_{i+2})$ that yield the same d(i,i+3) distance yet have a range of turn angles $\tau$. For larger d(i,i+3) distances the number of such quartets increases, as reflected in the wider colored regions. The general tendency observed is increased planarity of the turn ($\mu \to 0°$) leads to bigger turn angles, as one might intuitively expect.



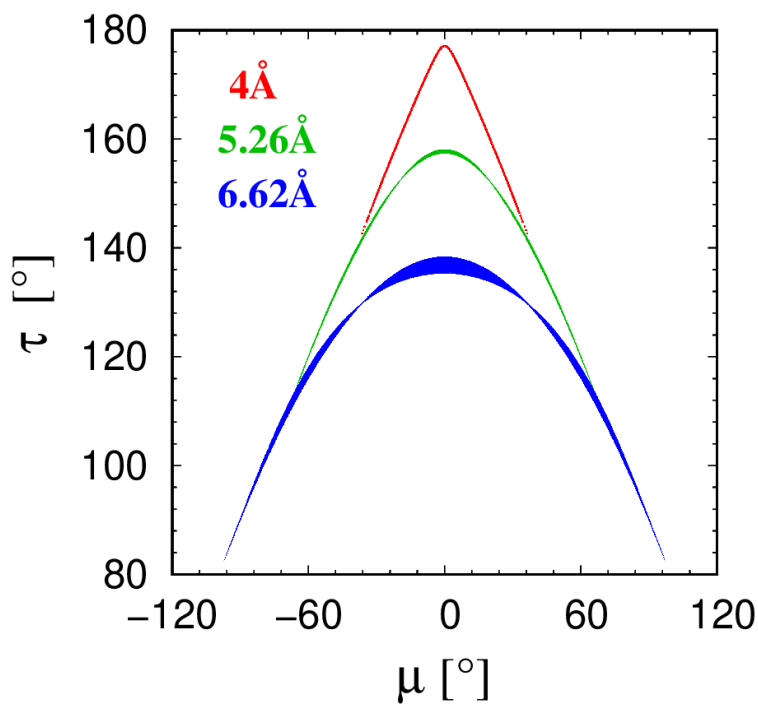

Figure 3: Cross plots of the turn angle τ versus the dihedral angle μ of quartets having a given fixed distance d(i,i+3) between the pair of beads (i,i+3).

## 2.2 Local protein contacts: symmetric poking, one-way poking and no poking

We have shown recently [30,31] that the notion of poking pairwise interactions is a simple yet powerful concept allowing for the development of a unifying framework to understand the formation of both secondary and tertiary protein structure. Here we aim to show that a pairwise poking contact is a key



concept in the analysis of protein turns as well. Figure 4 illustrates an arrangement of two pieces of a discrete chain such that a pairwise poking contact is established between two $C_\alpha$ atoms, i and j, located at the positions $\mathbf{r_i}$ and $\mathbf{r_j}$, respectively. The point i is closer to j than its two neighbors along the chain, and likewise for j, so that the pieces of the chain effectively 'poke' towards each other:

$$d(i,j) < d(i,j-1) \qquad \text{Eq. (1a)}$$

$$d(i,j) < d(i,j+1) \qquad \text{Eq. (1b)}$$

$$d(i,j) < d(i-1,j) \qquad \text{Eq. (2a)}$$

$$d(i,j) < d(i+1,j) \qquad \text{Eq. (2b)}$$

The term 'poking contact' is symmetric in nature. However, one can introduce the concept of asymmetric or 'one-way' poking contacts as well. 'One-way' poking contacts correspond to just the top two (Eq. (1a) and Eq. (1b)) *or* the bottom two (Eq. (2a) and Eq. (2b)) holding. Finally, there is 'no-poking' contact when at least one of the inequalities Eq. (1a) or Eq. (1b) is not satisfied *and* at the same time at least one of the two bottom inequalities Eq. (2a) or Eq. (2b), is also not satisfied.



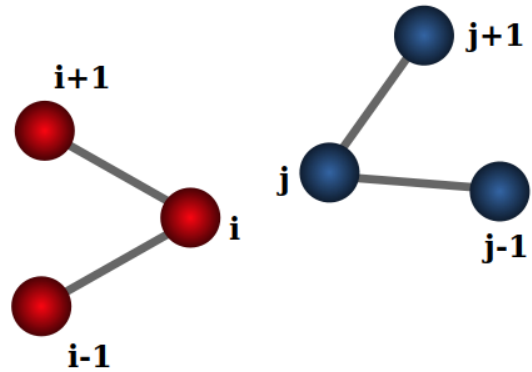

Figure 4: Schematic representation of the relative position of two pieces of a chain (i-1,i,i+1) and (j-1,j,j+1) each comprising of three consecutive beads that result in a poking pairwise contact between i and j. i is closer to j than to the two neighbors of j and j is closer to i than to the two neighbors of i.

The final concept we introduce here is the notion of a 'good' contact when a contact is not poking but is close. A good contact is one that is on the verge of becoming a symmetric poking contact. If a decrease of the distance between (i,j) within 0.263Å (or 10% of Δ) is sufficient to make (i,j) a genuine poking contact, we denote it as a good contact. Otherwise, we call it a bad



contact to indicate that it is not close to being a genuine symmetric poking contact.

## 2.3 Data set of native structures of globular proteins

Our protein data set consists of 4,391 globular protein structures from the PDB, a subset of Richardsons' Top 8000 set [33] of high-resolution, quality-filtered protein chains (resolution < 2Å, 70% PDB homology level), that we further distilled out to exclude structures with missing backbone atoms, as well as amyloid-like structures. Program DSSP (CMBI version 2.0) [34] was used to identify the presence of hydrogen bonds, as well as to determine for each protein residue if it was located within an α-helical, or β-strand environment or elsewhere.

## 2.4 Protein quartets analysis

In this work we are after an objective definition of protein turns, the smallest portions of a protein chain that can effectively change the chain direction. To this end, we have analyzed all 940, 538 quartets present in our data set of more than 4,000 globular proteins. We used DSSP (CMBI version 2.0) [34] to



divide all these quartets into three disjunctive groups: purely α-helical quartets (in which all four beads were assigned the flag 'H' by DSSP), purely extended β-strand quartets (for which all four beads were assigned the flag 'E' by DSSP) and all 'other' quartets. In this way we have identified 226,478 α-helical quartets, 98,648 β-strand quartets and 615,412 'other' quartets. The largest class of 'other' quartets includes those that are in the beginnings and the endings of the secondary structure elements, as well as those that are in protein loops.

## 2.5 Identification of structural templates in protein turns

We will demonstrate later in the Results section that the purest turns correspond to isolated poking contacts (meant in the full symmetric sense). We first consider all 21,571 isolated (i,i+3) poking contacts located in 59,464 loops of 4,391 globular protein chains of our data set. We further ensured that all quartets distilled in this manner have standard (or 'long') bonds (whose values are equal to 3.81Å within ±2% [32]). We note that isolated (i,i+3) poking contacts with at least one 'short' bond length of ~3Å are very few in number, only 228 of them, and will be analyzed separately. The occurrence of



these 'short' $C_\alpha - C_\alpha$ pseudo-bonds are due to the rare *cis*-configuration of the protein backbone (third Ramachandran angle being $\omega \approx 0°$, so that the neighboring $C_\alpha$ atoms along the chain are on the same side of the planar peptide bond), as opposed to the common *trans*-configuration (when $\omega \approx 180°$, with the neighboring $C_\alpha$ atoms along the chain on the opposite sides of the planar peptide bond, and thus further away) [35].

In order to systematically assess the structural variety of protein turns, we have analyzed the distribution of 21,571 quartets all having standard bonds in the $(\theta_{i+1}, \theta_{i+2}, \mu)$ three-dimensional space. The distribution of the density of quartets in the three-dimensional space can be characterized by identifying significant local maxima to create structural templates. We separately treat 18,096 'right-handed' quartets $(\mu \geq 0°)$ and 3,475 'left-handed' quartets $(\mu < 0°)$, because it is not possible to superimpose an object on its mirror image by mere rotations and translations. The optimal binning of the three-dimensional $(\theta_{i+1}, \theta_{i+2}, \mu)$ structural space of quartets was found to be 2.5° for the two bond bending angles $\theta_{i+1}$ and $\theta_{i+2}$ and 5° in the dihedral angle $\mu$. Three templates were identified for the right-handed class populated with 394, 132 and 92 quartets (corresponding to Regions I, III and II, respectively, see the



Results section). For the left-handed class, two local maxima were identified containing 76 and 56 quartets (corresponding to Regions V and IV, respectively). We then worked out the geometries of the five structural templates using the mean values of three angles $\theta_{i+1}$, $\theta_{i+2}$, and $\mu$ in the bins (see Table II in the Results section).

## 2.6 Mapping protein turns

Each of the 21,571 isolated (i,i+3) poking contacts with standard bond lengths was assigned to one of the five basins associated with the five structural templates. For a given protein quartet, the actual values of the three bond lengths (i,i+1), (i+1,i+2) and (i+2,i+3) were measured and five templates were constructed with those values of the three bond lengths, as well as the values of the three mean angles $\theta_{i+1}$, $\theta_{i+2}$, and $\mu$ calculated for a given template in a manner described above. Then a given protein quartet and each template geometry were superimposed and the value of the root-mean-square deviation (RMSD) was calculated using the Visual Molecular Dynamics (VMD) Software package [36]. Finally, by identifying the structural template to which a given protein quartet was closest to (i.e., having the smallest



RMSD), the protein quartet was assigned as belonging to that template basin. In this way, the map of the protein turns was determined.

## 3 Results

## 3.1 Theoretical considerations

In our recent work [30,31], we introduced a concept of poking pairwise interactions for a discrete chain, a simple yet powerful concept that allowed understanding of the formation of both secondary and tertiary protein structure. The geometry of α-helices is found to be the unique arrangement in which local $(i,i+3)$ poking contacts with special geometrical constraints are accommodated in a repetitive manner. On the other hand, β-sheets are geometries that have pairwise poking non-local $(i,j)$ contacts with $(j-i) \geq 4$ along a pair of coupled β-strands. It was argued that pairwise poking interactions also govern, in a harmonious and compatible manner, tertiary structure formation, but they are non-repetitive and weaker with no special geometrical constraints.



Here we will demonstrate that protein turns, the most prevalent non-repetitive structural element in proteins, are also instances of local and isolated (non-repetitive) poking pairwise contacts for which the geometrical constraints are partially relaxed. Indeed, it is intuitively obvious that when tight turns are not constrained to periodically repeat as in protein α-helices, they can become more planar and offer a simple rationalization for their observed structural variety and versatility. That this is indeed the case can be seen in the Supplementary Information (Figure SI 1), which shows a detailed comparison between the constraints required for repetitive (completely embedded) and completely isolated poking contacts in proteins.

## 3.2 What is an objective definition of protein turns?

We ask the basic question: which type of a consecutive quartet of beads, the minimal portion of the local structure that effectively can turn the chain, can be thought of as belonging to the 'true turn' category? Can one come up with an objective definition of a protein turn? George Rose [28] has pointed out that a β-turn quartet is solvent accessible, it often has a hydrogen bond



between residues 1 and 4 (or equivalently i and i+3) and lies at the protein surface because of the inability of the two middle residues to satisfy their backbone polar groups within the turn without a steric clash. Unlike the building blocks of helices and sheets, β-turns are neither iso-directional nor completely solvent-shielded. Turns can form autonomously and potentially initiate the assembly of scaffold elements by bringing them together with a desired relative placement. Here we observe that the formation of a small piece of a helix is necessarily the only way to nucleate a helix [37-38]. In our way of thinking, this is an isolated local (i,i+3) poking contact that, similarly to within helices, is a tight turn but without the baggage of geometrical constraints necessitated by a properly repeating helix.  Such pieces are candidates for growing into a fully formed helix. We will argue that when such growth is thwarted for one reason or another, they can also serve, in a flexible manner by tuning their degree of planarity, as one of the ubiquitous classes of turns that allow the protein to become overall compact and break the iso-directionality of a strand or a helix.

We have studied all 940, 538 quartets that are present in our data set of more than 4,000 globular proteins. We use the DSSP program [34] to divide



all quartets into three disjunctive groups: purely α-helical quartets, purely extended β-strand quartets, and 'other' quartets, as explained in the Materials and Methods section. Figure 5 shows the (τ-μ) cross-plot for these three groups of quartets (in Panels (A), (B) and (C)). Panel (D) shows the same cross-plot but this time for *all* the quartets, which have an isolated poking contact of (i,i+3). Figure 6 shows the corresponding frequency distributions of the quartet turn angles τ. As expected, quartets with an isolated (i,i+3) poking contact do indeed display large turn angles τ and they are primarily located in the 'other' class of quartets, that is mainly in protein loops [17,23].



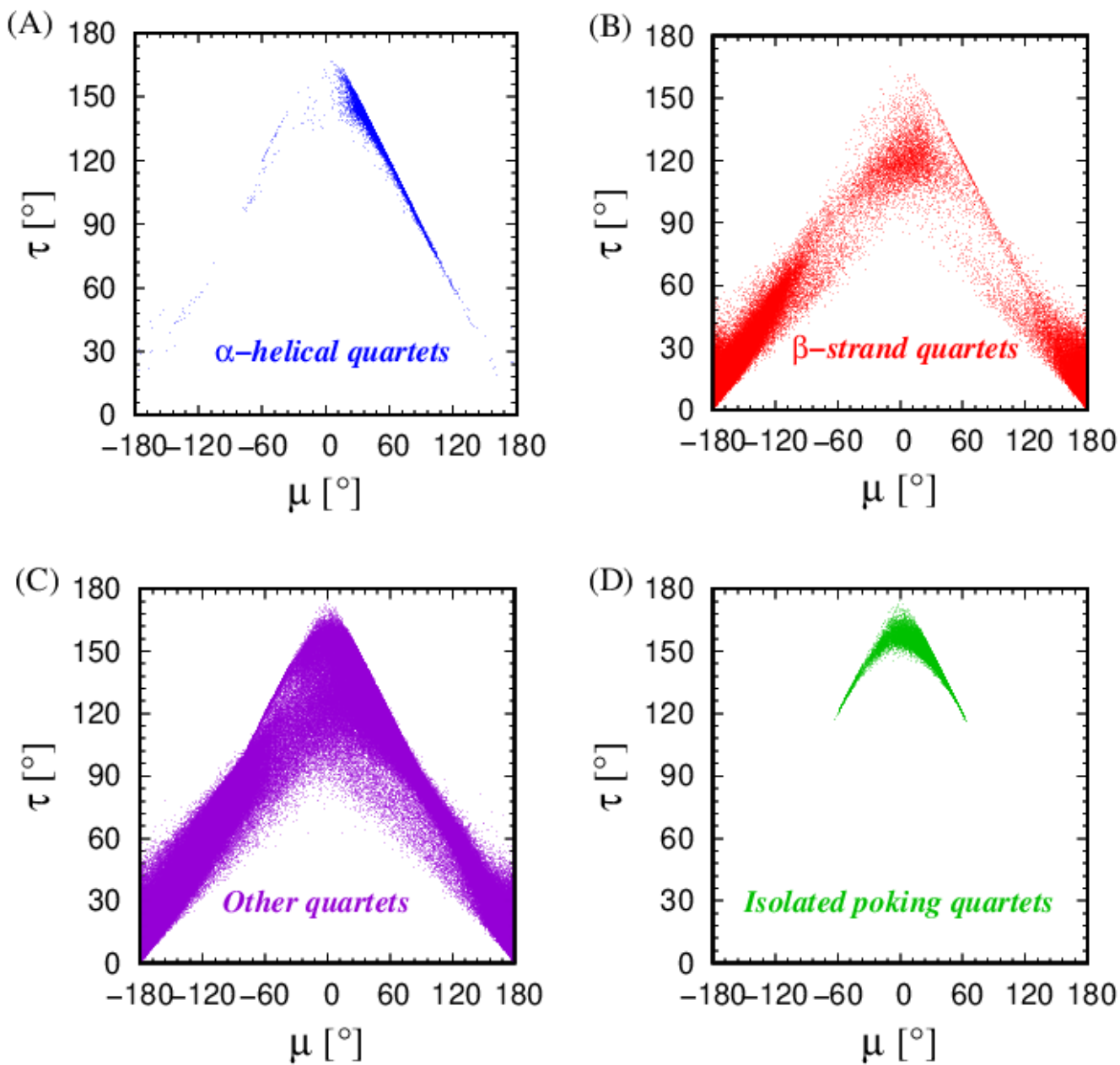

Figure 5: Cross-plots of the turn angles τ vs dihedral angles μ for the four classes of protein quartets: Panel (A) shows the frequency distribution of the turn angles τ in the 226,478 α-helical quartets in our data set (blue histogram), Panel (B) shows the frequency distribution



of the turn angles $\tau$ in the 98,648 β-strand quartets in our data set (red histogram),  Panel (C) shows the frequency distribution of the turn angles $\tau$ in the 615,412 'other' quartets in our data set (purple histogram), and finally in Panel (D) the frequency distribution of the turn angles $\tau$ in the 25,618 isolated poking quartets in our data set is shown (green histogram). We note that more than 95% of the (i,i+3) isolated poking contacts in our data set (24,349 in all) are in the class of 'other' quartets. There are 1,158 (i,i+3) isolated poking contacts in α-helical quartets and a mere 111 in the β-strand quartets.



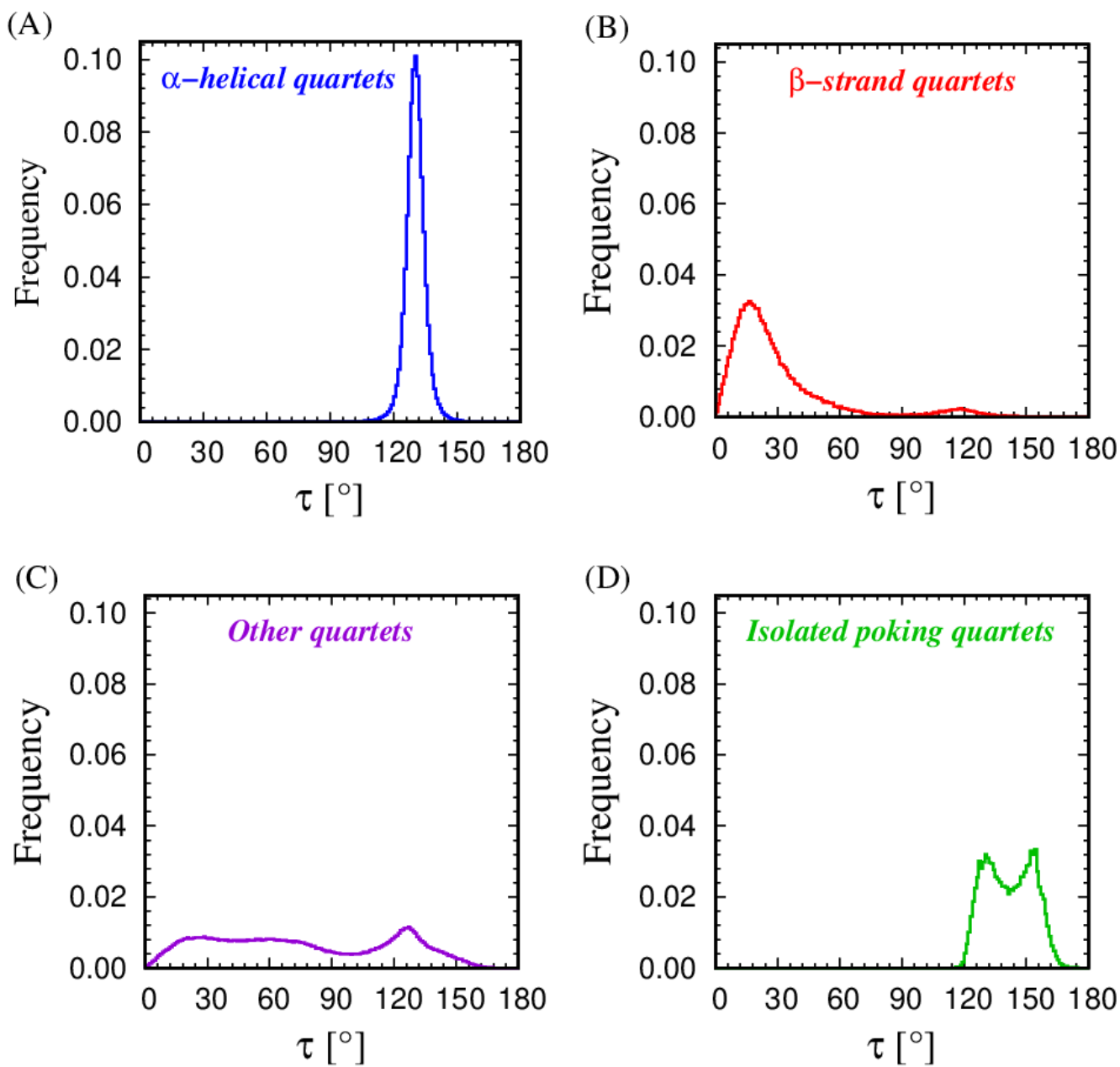

Figure 6: Distribution of the turn angles for the four classes of protein quartets: Panel (A) shows the frequency distribution of the turn angles $\tau$ in the 226,478 α-helical quartets in our data set (blue histogram),



Panel (B) shows the frequency distribution of the turn angles τ in the 98,648 β-strand quartets in our data set (red histogram), Panel (C) shows the frequency distribution of the turn angles τ in the 615,412 'other' quartets in our data set (purple histogram), and finally in Panel (D) the frequency distribution of the turn angles τ in the 25,618 isolated poking quartets in our data set is shown (green histogram).

We inspect the distance between quartet ends, d(i,i+3), and measure its frequency distributions across different quartet groups, and the result is shown in Figure 7. We see that helical quartets and quartets with isolated (i,i+3) poking contacts are similar (as expected) and have a d(i,i+3) distance around the value of the 'Kepler coin' diameter 2Δ [30,31]. The distances for the extended β-strand quartets are understandably the largest among all groups with the 'other' being a mixture of the other three behaviors [17,23].



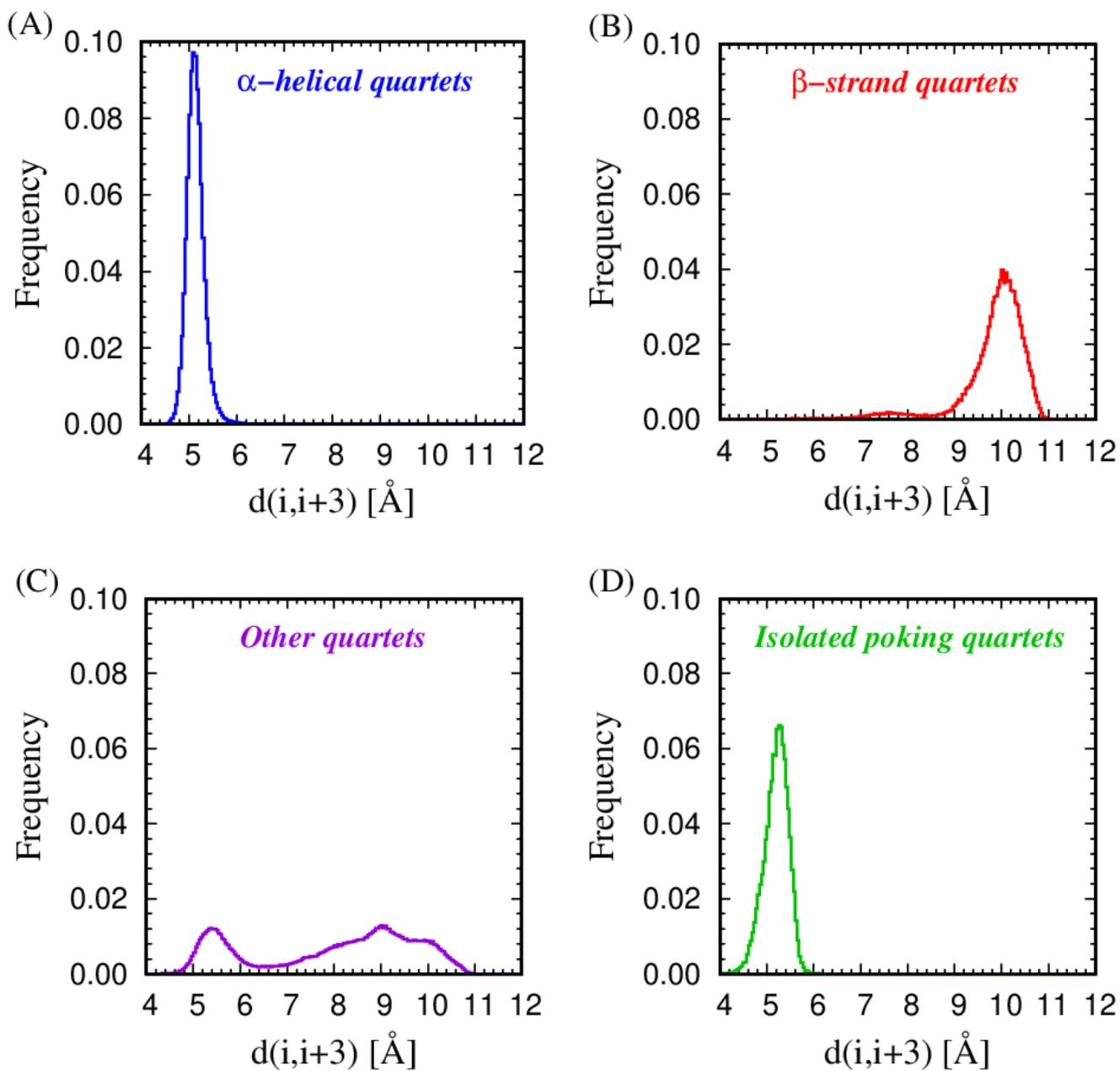

Figure 7: Same as in Figure 6, but for the distribution of frequencies of the (i,i+3) distances for the four classes of quartets.



The planarity of protein quartets is probed by analyzing the frequency distributions of their dihedral angles μ, shown in Figure 8. The local and repetitive helical quartets are decisively non-planar with dihedral angles around ~50°, while extended strand quartets are close to planar with |μ| ~ 180°. The planar quartets with μ~0° are located in the quartets with isolated (i,i+3) poking contacts and are predominantly a subset of the group of 'other' quartets. We have verified that the small weight around μ~0° even in the group of β-strand quartets is present because the 'E' flag assigned by DSSP [34] sometimes includes hairpin loops.

Figure 9 shows the distribution of frequencies of the bond bending angles $\theta_{i+1}$ and $\theta_{i+2}$ in the four classes of quartets analyzed together. We see that with the exception of extended β-strand quartets, all the rest contain a significant amount of tight helix-like bond bending angles of around ~90°. Our principal finding from the rudimentary analysis is that quartets with isolated (i,i+3) poking contact are in fact plausible turn candidates and are predominantly located in protein loops [17,23].



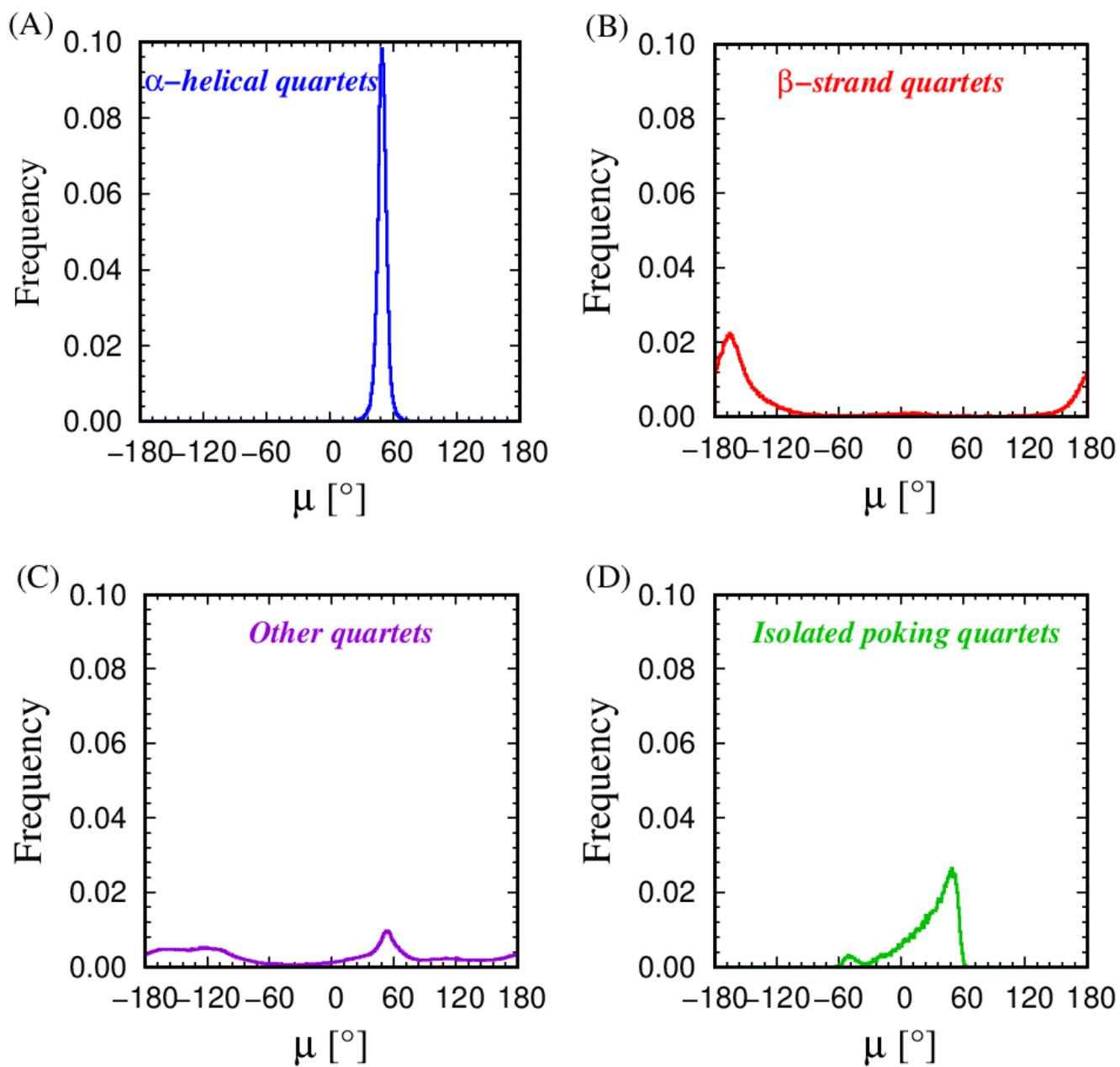

Figure 8: Same as in Figure 7, but for the distribution of frequencies of the dihedral angles μ for the four classes of quartets.



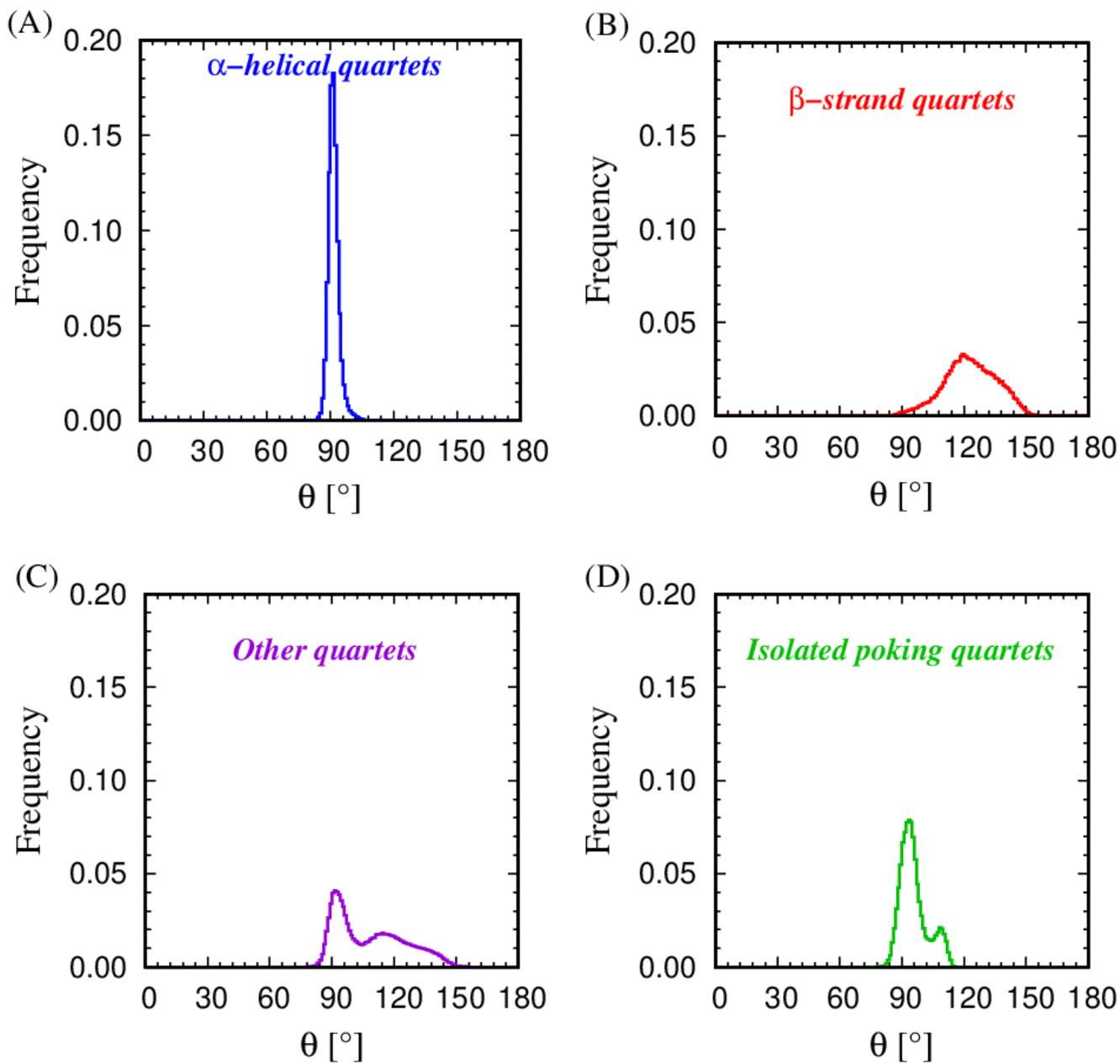

Figure 9: Same as in Figure 8, but for the distribution of frequencies of the bond bending angles in the four classes of quartets. There are two bond-bending angles in a quartet and they are both treated together in this analysis.



We now proceed to study the isolated (i,i+3) poking contacts located within protein loops [17,23], thereby excluding those that are in the beginning and endings of helices and sheets. We define a protein loop as a segment of the protein that connects two elements of protein structure along the chain, namely α-α, β-β, α-β, or β-α. Approximately 85% of isolated (i,i+3) poking contacts are located in protein loops. The rest occur at the very beginning or ending of a protein, as well at the edges of secondary structure elements. In total, there are 59,464 loops in our proteins out of which 8,675 (14.6%) have length of four (implying that there are two 'O' (beads other than 'E' or 'H', located in strands and helices, respectively) beads at positions 2 and 3). There are 7,557 loops of length five (12.7%), 8,395 loops of length six (14.1%), 7,829 loops of length 7 (13.2%), 5,379 loops of length eight (9.0%), 4,071 loops of length nine (6.8%), and 3,245 loops of length 10 (5.5%). Thus, we find that the number of loops decreases approximately monotonically with the length of the loop, a trend noted previously in Ref. [17,23].



We are now ready to take the next step in our search for the objective definition of protein turns. To this end, we discuss the detailed topology of the (i,i+3) contacts in protein quartets located in protein loops, in terms of whether they are symmetric poking contacts and, if not, whether they are good or bad (as per the definition introduced in Materials and Methods Section 2.2). Figure 10 shows the frequency distributions of the turn angles τ in protein quartets in protein loops, divided in different classes according to their topological characteristic. Panel (A) of Figure 10 underscores the sharp difference in the turn angle distributions for good and bad contacts with no poking at all -- the good ones decisively have a larger turn angle. Panel (B) of the same Figure shows that even for 'one-way' poking contacts, the good ones have a larger weight for bigger turn angles. Finally, Panel (C) shows that among symmetric poking contacts, the isolated ones turn the chain more sharply than the non-isolated ones.



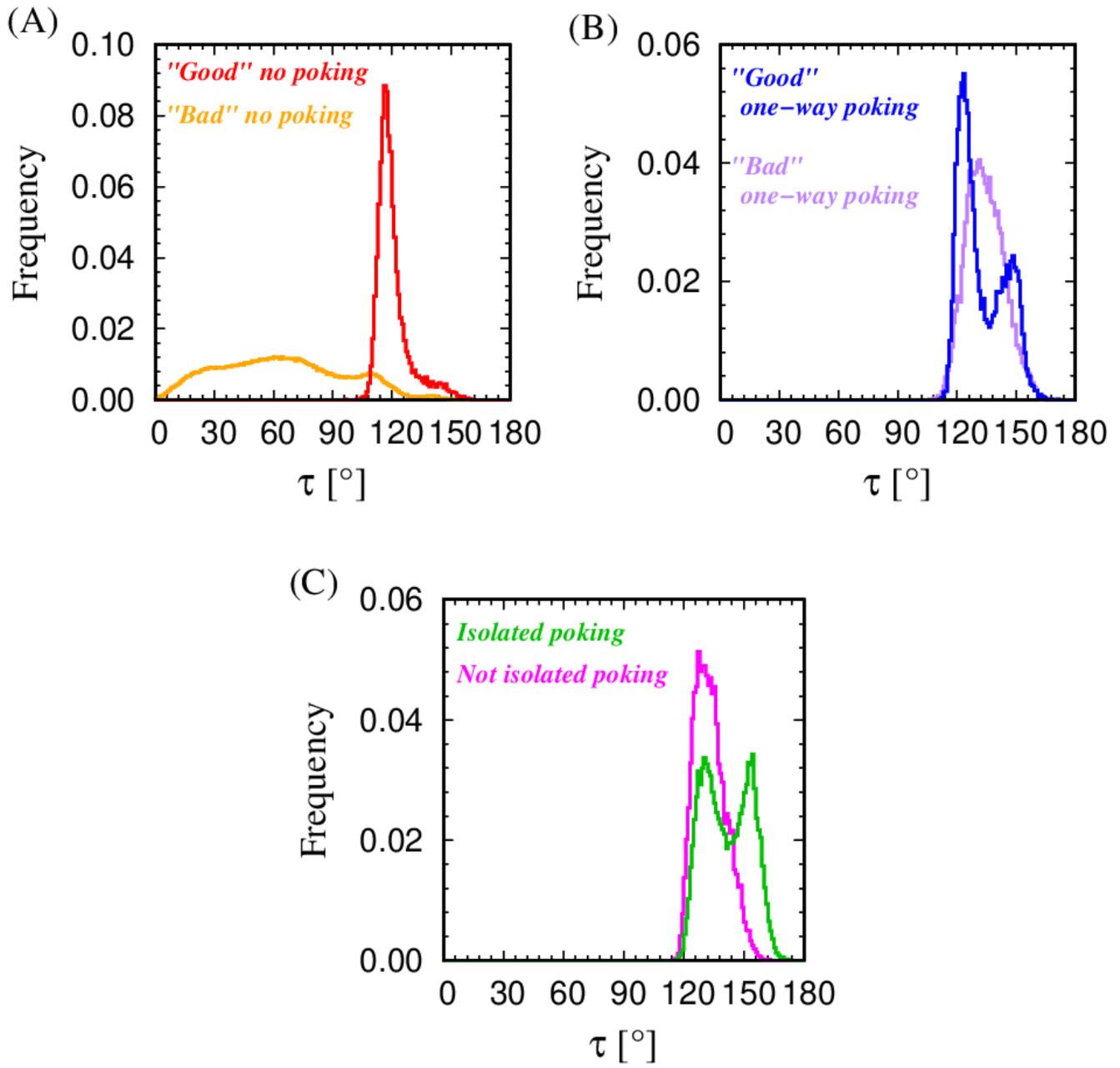

Figure 10: Turn angle distributions. Panel (A) shows the frequency

distribution of the turn angle τ for the classes of 11,954 'good' and



282,813 'bad' (i,i+3) contacts in protein loops with no poking at all (in red and orange). The contact is classified as 'good' if its effective 'distance' from the symmetric poking condition is less than 10% of Δ (0.263Å). Panel (B) shows the frequency distribution of the turn angle τ for the class of 17,721 'good' and 11,954 'bad' (i,i+3) contacts in protein loops that have a one-way poking contact (in blue and violet respectively). Finally, Panel (C) shows the frequency distribution of the turn angle τ for the 21,571 (i,i+3) isolated poking contacts (fully symmetric) as well as 6,886 (i,i+3) isolated poking contacts that are not fully isolated ('embedded from one side') in protein loops (in green and magenta).

Table I shows the number and percentages of turns belong to the 'topological' classes with respect to their (i,i+3) 'poking' status. In this way, we find that the purest class of sharp protein turns is indeed the one making isolated (i,i+3) poking contacts.



| Type of (i,i+3) contacts | All | Loops |
|:---:|:---:|:---:|
| | $\tau > 160°$ | $\tau > 160°$ |
| Isolated poking | 1,042 | 959 |
| Not isolated poking | 3 | 1 |
| 'Good' one-way poking | 96 | 74 |
| 'Bad' one-way poking | 85 | 53 |
| 'Good' no poking | 34 | 7 |
| 'Bad' no poking | 14 | 2 |
| | Total: 1,274 | Total: 1,096 |

Table I: Number of quartets with large turn angles, $\tau > 160°$, divided into classes of (i,i+3) contacts. 'All' considered all 940,538 quartets in all our proteins whereas 'Loops' was restricted to 357,989 protein quartets just within loops (see Materials and Methods Section 2.4 for a description). The clear result is that large turn angles are predominantly associated with isolated poking contacts, the nuclei for the putative formation of protein helices.



### 3.3 Geometrical classes of protein turns

Figure 11 depicts various cross-plots of geometrical attributes of protein turns defined as quartets of beads with (i,i+3) isolated poking contact (in Panels (A) and (B)), and contrast them with the geometrical attributes of quartets in the class of embedded (repetitive) (i,i+3) poking contacts (Panels (C) and (D)). The results of theoretical estimates are also shown alongside (see Figure caption for a description). There is a clear contrast between the two classes: quartets with 'isolated' (i,i+3) poking contacts explore a much wider region of phase space and are much less constrained than the embedded helical contacts, which are restricted by the need for repetition.



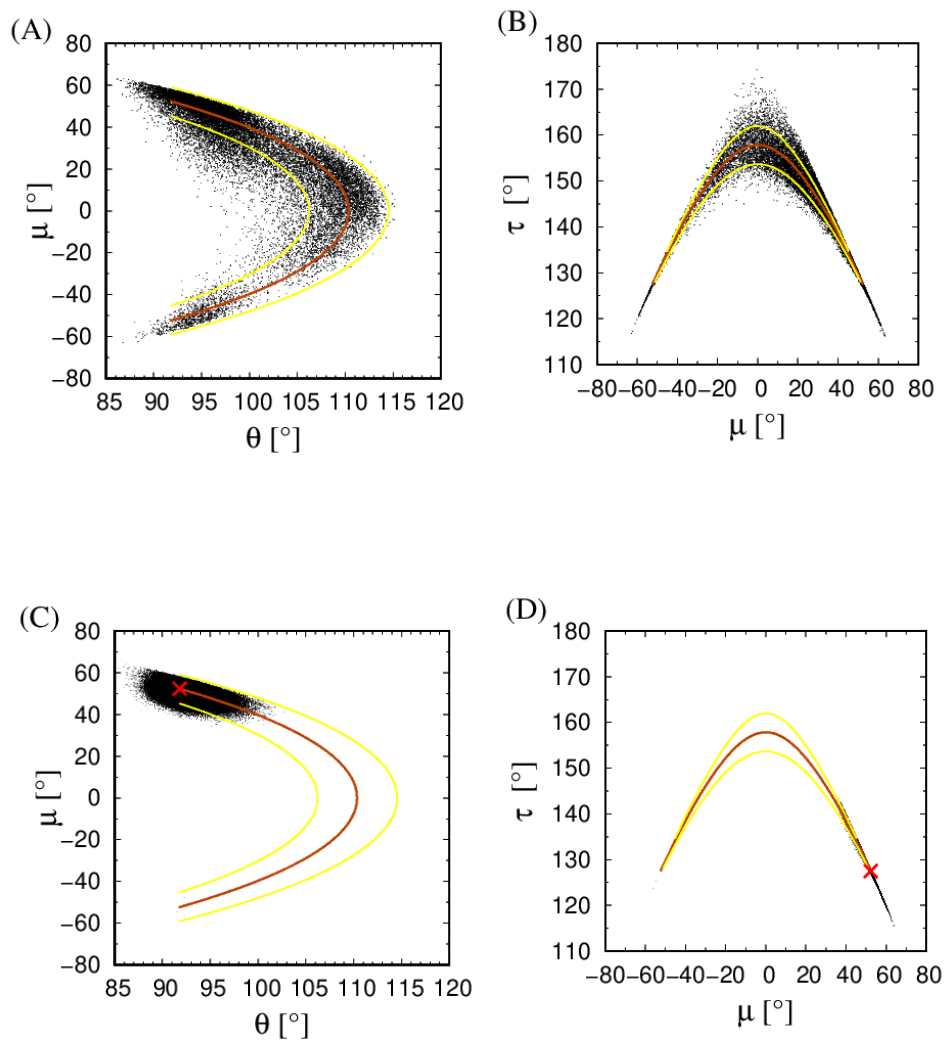

Figure 11: Panel (A) shows the (θ,μ) cross plot of the full set of 21,571 isolated (i,i+3) symmetric poking contacts located in the 59,464 loops of 4,391 globular protein chains of our data set, along with the



theoretical prediction curve (in dark orange color) for the dependence of the dihedral angle μ of the turn and the larger of the two bond bending angles θ, [that is θ = $θ_1$, if $θ_1 > θ_2$, and θ = $θ_2$, if $θ_2 > θ_1$]. Panel (B) shows the (μ,τ) cross plot of the full set of 21,571 isolated (i,i+3) symmetric poking contacts located in the 59,464 loops of 4,391 globular protein chains of our data set. Panels (C) and (D) depict the same quantities as (A) and (B), but for the 176,711 embedded (i,i+3) symmetric poking helical contacts in 4,391 globular protein chains of our data set. The red X symbols in both panels (C) and (D) denote the corresponding values for the Kepler helix. The theoretical predictions are derived from the condition that the distance between the first and the fourth bead in the turn (i,i+3) equals 2Δ = 5.26Å (the ideal distance for a Kepler helix), when one bond bending angle is fixed to correspond to the turn angle of a Kepler helix, 91.8°. The yellow curves in both panels are theoretical predictions when a ±5% difference is allowed in the theoretical 2Δ distance d(i,i+3).



As explained in Sections 2.5 and 2.6 of Materials and Methods, a quartet with standard bonds is assigned one of five representative structural templates. The five-colored map obtained in this way is shown in Panels (A) and (B) of Figure 12. Panels (C) and (D) show the case of quartets with embedded (i,i+3) poking contacts (helical turns) that are described with just one local maxima and one structural template. Region I (green) of protein turns corresponds to the unrelaxed helical turn. Figure 13 (A) shows the unique geometrical template for quartets containing embedded (i,i+3) poking contacts and correspond to the black X symbols in Panels (C) and (D) of Figure 12. The five structural templates of protein turns are shown in Panels (B-F) of Figure 13.

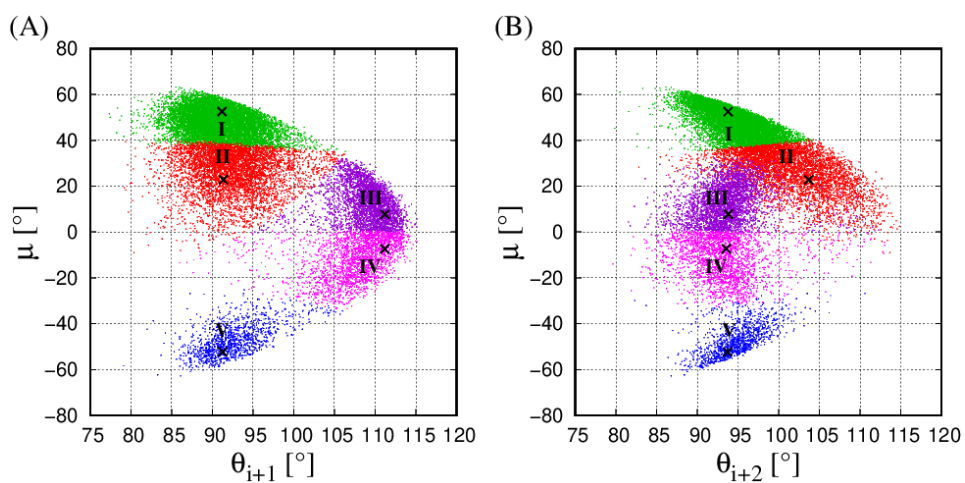



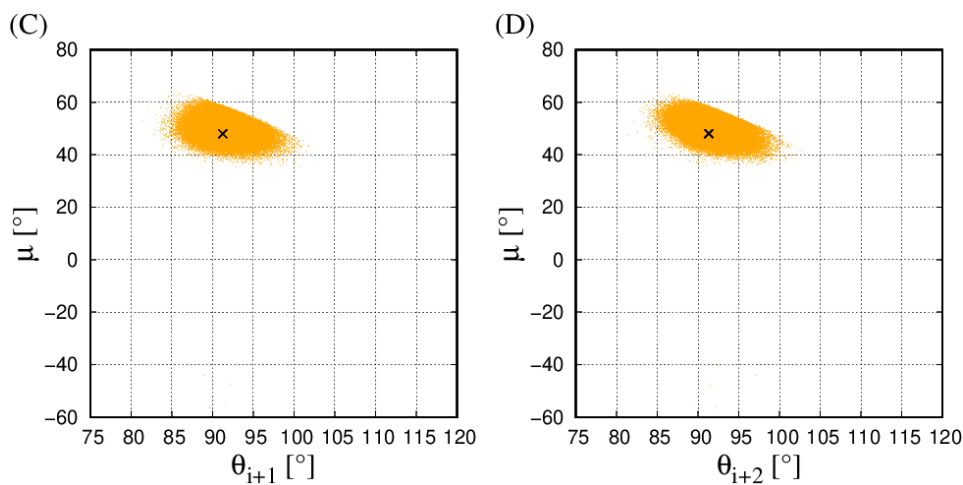

Figure 12: Panel (A) shows the map of the two dimensional projection in the ($\theta_{i+1}$, μ) plane of the full set of 21,571 isolated (i,i+3) symmetric poking contacts located in the 59,464 loops of 4,391 globular protein chains of our data set. The black X symbols denote the locations of the five local maxima in the three-dimensional space of the three angles ($\theta_{i+1}$, $\theta_{i+2}$, μ) that have the largest density of points with the binning of 2.5°,2.5°, and 5° in the three angles, respectively. Each of the 21,571 isolated (i,i+3) symmetric poking contacts is assigned to one of the five templates depicted in Panels (B)-(F) of Figure 13, based on the criterion of the smallest RMSD value from the winning template. The basin represented with Region I is found to contain



9,604 quartets (green points), Region II is found to contain 5,271 quartets (red points), Region III is found to contain 3,237 quartets (purple points), Region IV is found to contain 2,298 quartets (magenta points), and finally Region V is found to contain 1,161 quartets (blue points).  Panel (B) shows the cartography in a complementary two-dimensional plot in the ($\theta_{i+2}$, $\mu$) plane. Panels (C) and (D) depict the data for 176,711 embedded (i,i+3) symmetric poking contacts in 4,391 globular protein chains of our data set (points in orange color).  The black X symbols in Panels (C) and (D) denote the locations of the local maxima in the three-dimensional space of the three angles ($\theta_{i+1}$, $\theta_{i+2}$, $\mu$) with the binning of 2.5°,2.5°, and 5° in the three angles, respectively.



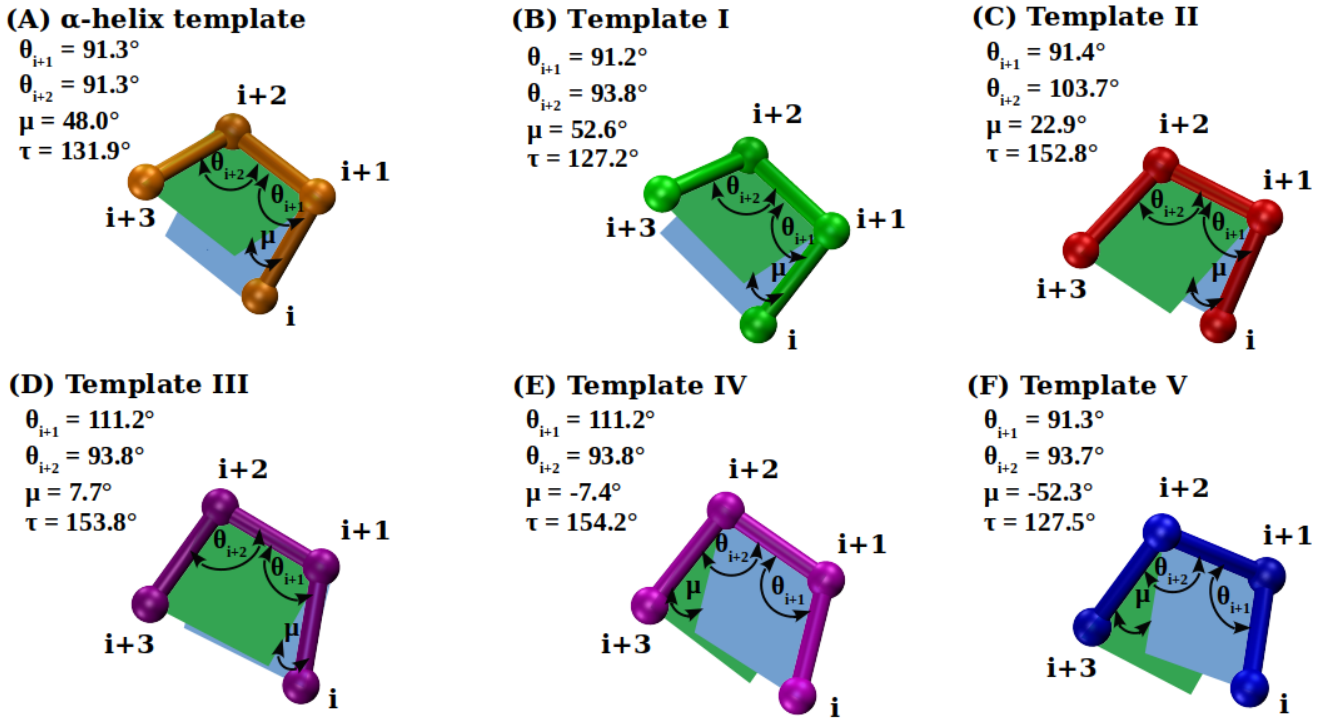

Figure 13: Panels (A) shows the single structural template for the embedded symmetric (i,i+3) poking contact characteristic of protein α-helices. Panels (B-F) show five structural templates determined for the local turns with isolated symmetric (i,i+3) poking contacts. The spheres represent the locations of four consecutive $C_\alpha$ atoms. The values of the bond bending angles $\theta_{i+1}$, $\theta_{i+2}$, dihedral angle $\mu$, and the turn angle $\tau$, for each of the six templates reported in the figure correspond to the mean values calculated over the three-dimensional bin associated with each of the identified five local maxima. The three bond lengths are all drawn with length b of 3.81Å. The planes



containing points (i,i+1,i+2) are shown in blue, while the planes containing points (i+1,i+2,i+3) are shown in green. When the green plane is above the blue plane, the quartet of points is right-handed (it has a positive value of the dihedral angle μ), whereas when the blue plane is above the green plane the quartet of points is left-handed (it has a negative value of the dihedral angle μ).

Figures 14 and 15 show the $(\theta,\mu)$ and $(\mu,\tau)$ cross plots, respectively, of the five structural templates of protein turns alongside the theoretical estimates. Interestingly, most turn geometries lie within a ±5% tolerance around the curve corresponding to the theoretical distance 2Δ between the quartet ends, d(i,i+3). The exception to this behavior is in Region II (red points) that represents a more planar version of the helical turn having a smaller value of the end-to-end quartet distance of 5.07Å (see Table II). This turn type has a lower percentage of hydrogen bonds [24] associated with its ends.

Figures 16 shows Ramachandran plots $(\varphi_{i+1}, \Psi_{i+1})$ and $(\varphi_{i+2}, \Psi_{i+2})$ [29] for quartets with the embedded (i,i+3) contacts, with black X symbols



showing their mean values [3]. Figure 17, Panels (A-E) shows Ramachandran plots ($\varphi_{i+1}$, $\Psi_{i+1}$) and ($\varphi_{i+2}$, $\Psi_{i+2}$) for the five geometrical regions identified in protein turns, along with the black X symbols, at the positions of the conventional mean values for Type I and Type II turns, along with their mirror images (Types I' and II') predicted by Ventakachalam [3].

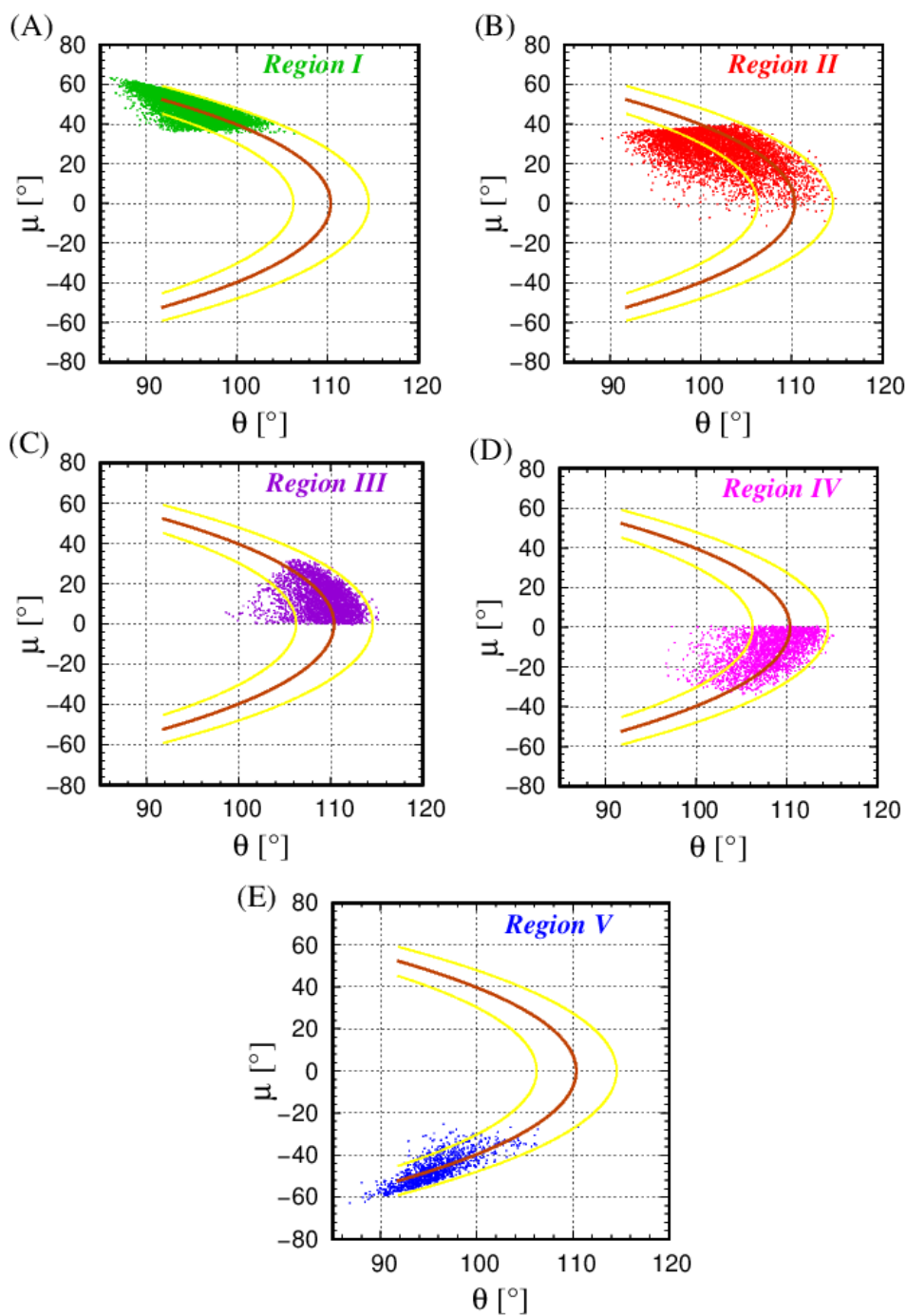

Figure 14: Panels (A-E) show separate (θ,μ) cross plots for the five regions for 21,571 total isolated (i,i+3) symmetric poking contacts located in the 59,464 loops of 4,391 globular protein chains of our data set, along with the theoretical prediction curve (in dark orange color) for the dependence of the dihedral angle μ of the turn and the larger of the two bond bending angles θ, [that is θ = $θ_1$, if $θ_1$> $θ_2$, and θ = $θ_2$, if $θ_2$> $θ_1$]. The theoretical predictions are derived from the condition that the distance between the first and the fourth bead in the turn (i,i+3) equals 2Δ = 5.26Å (the ideal distance for a Kepler helix), when one bond bending angle is fixed to correspond to the tight turn angle of a Kepler helix, 91.8°. The yellow curves in both panels are theoretical predictions when a ±5% difference is allowed in the theoretical 2Δ distance d(i,i+3).



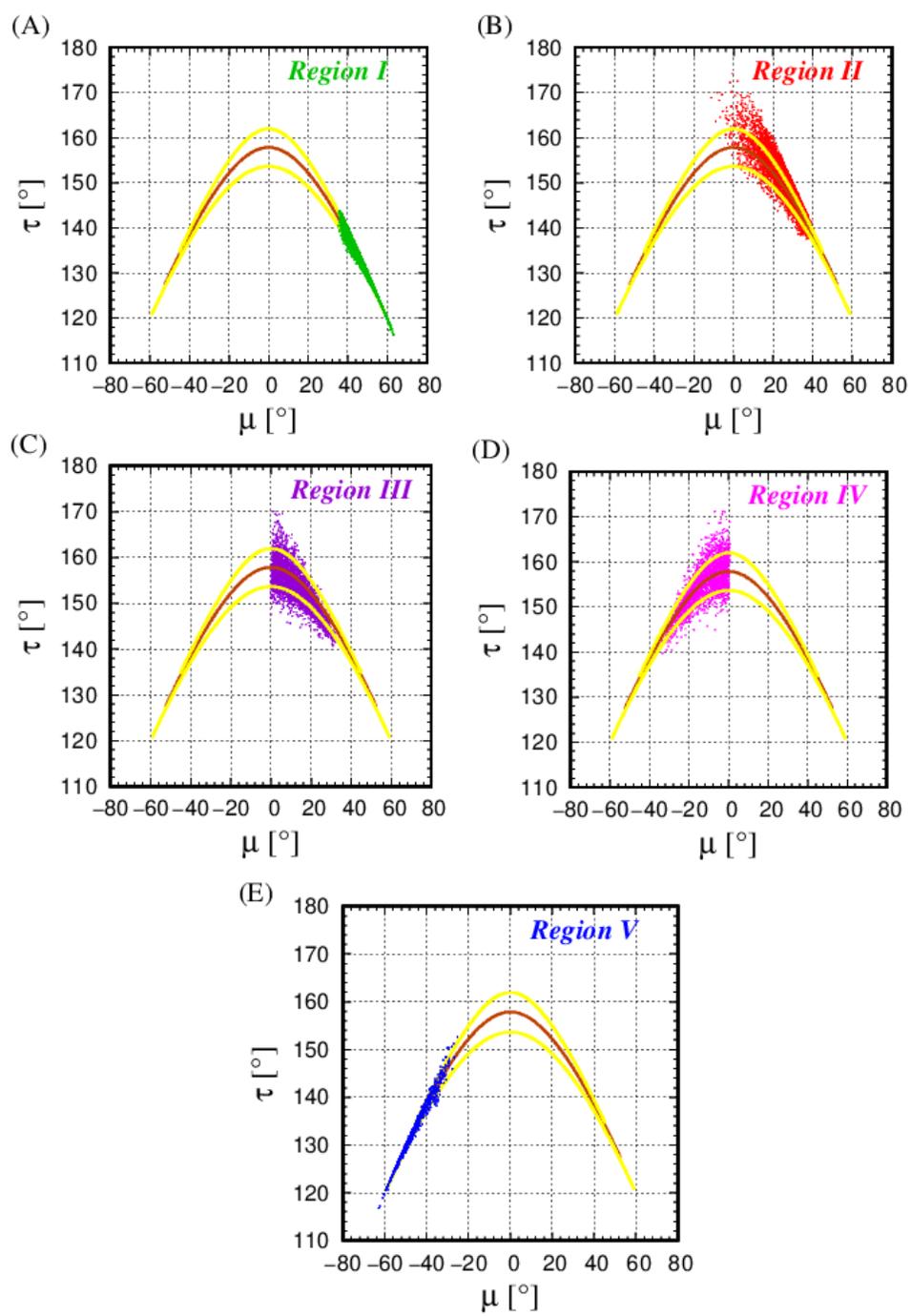



Figure 15: Panels (A-E) show the (μ,τ) cross plots for each of the five regions of the full set of 21,571 isolated (i,i+3) symmetric poking contacts located in 59,464 loops of 4,391 globular protein chains of our data set, along with the theoretical estimate curves (in dark orange color).

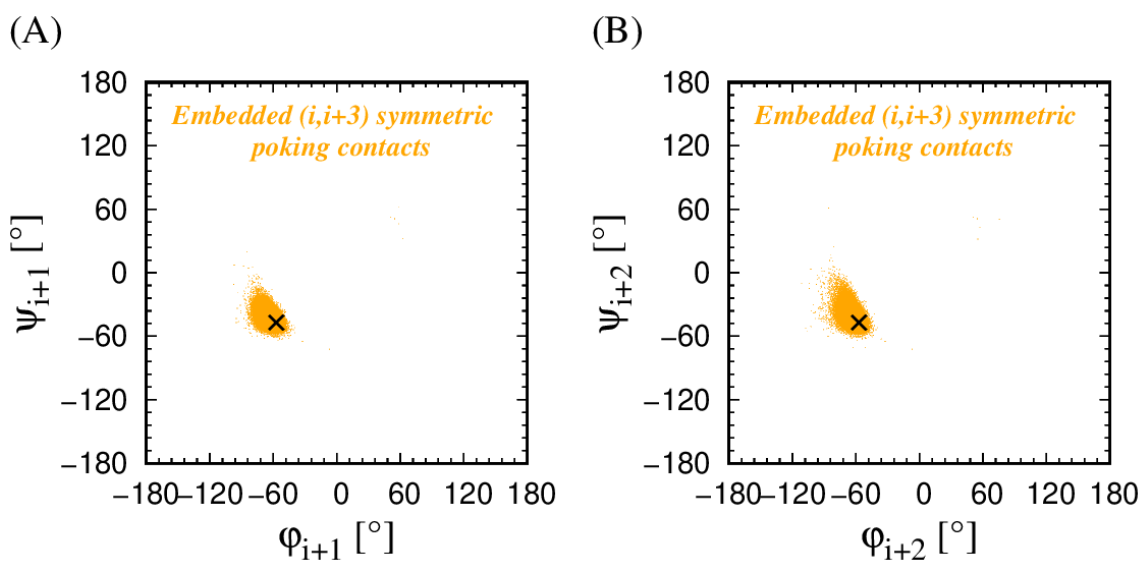

Figure 16: Ramachandran plots [29] for the beads (i+1) and (i+2) in the middle of the quartet (i,i+1,i+2,i+3), when (i,i+3) is embedded symmetric poking contact, ($\varphi_{i+1}$, $\Psi_{i+1}$) and ($\varphi_{i+2}$, $\Psi_{i+2}$), are shown in panels (A) and (B), respectively. The black X symbols indicate typical average values for protein α-helices indicated in the literature.



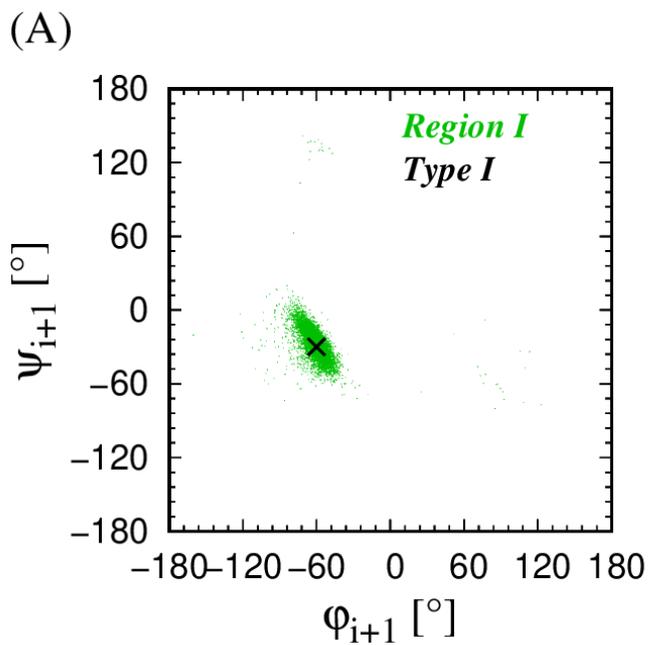

(A)

*Region I*
*Type I*

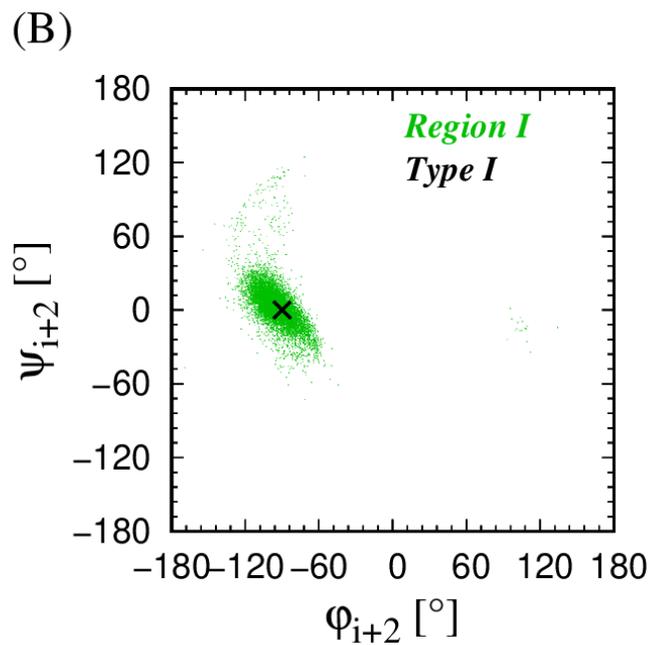

(B)

*Region I*
*Type I*

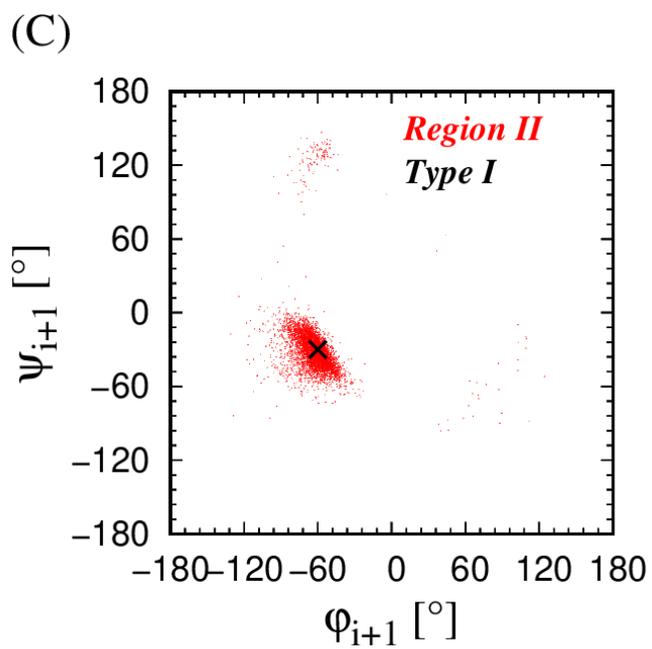

(C)

*Region II*
*Type I*

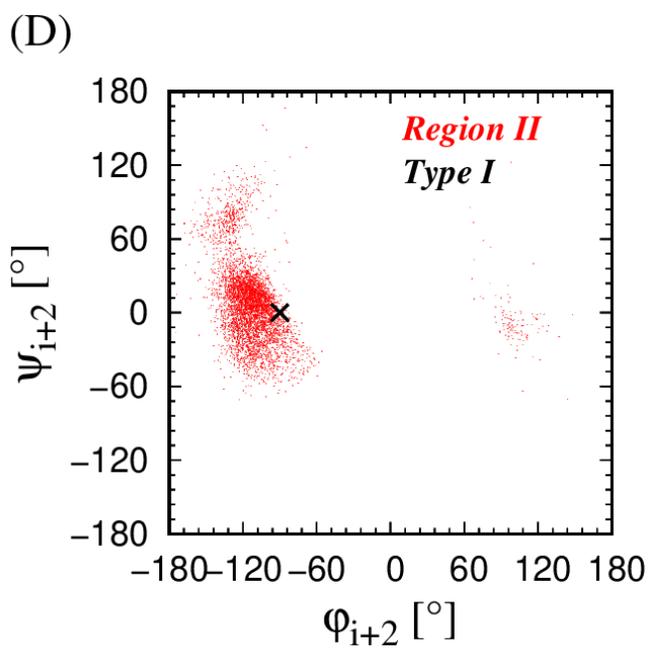

(D)

*Region II*
*Type I*



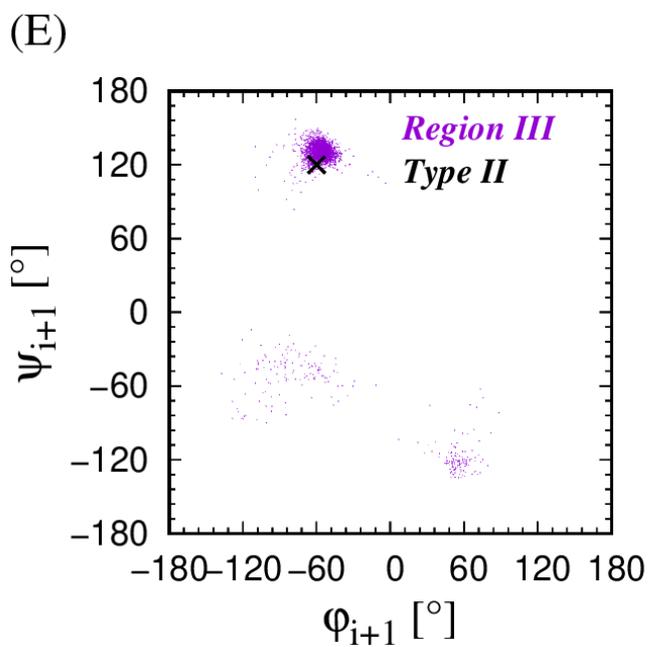

(E)

*Region III*
*Type II*

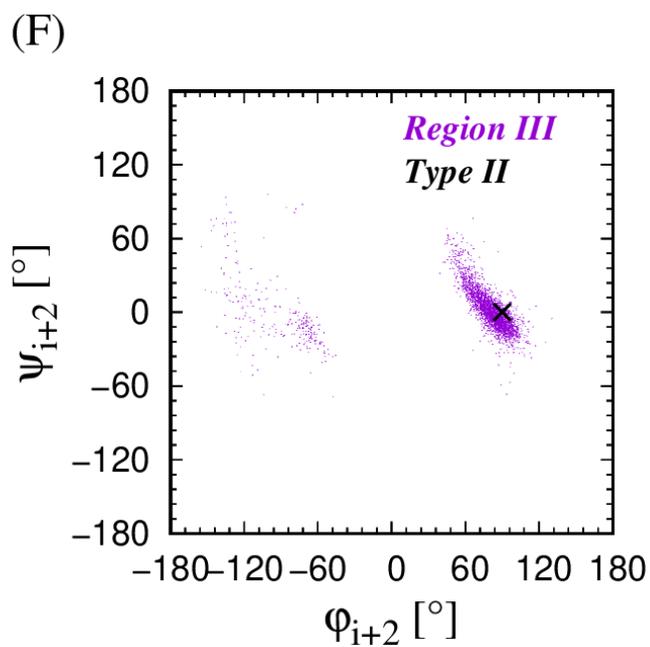

(F)

*Region III*
*Type II*

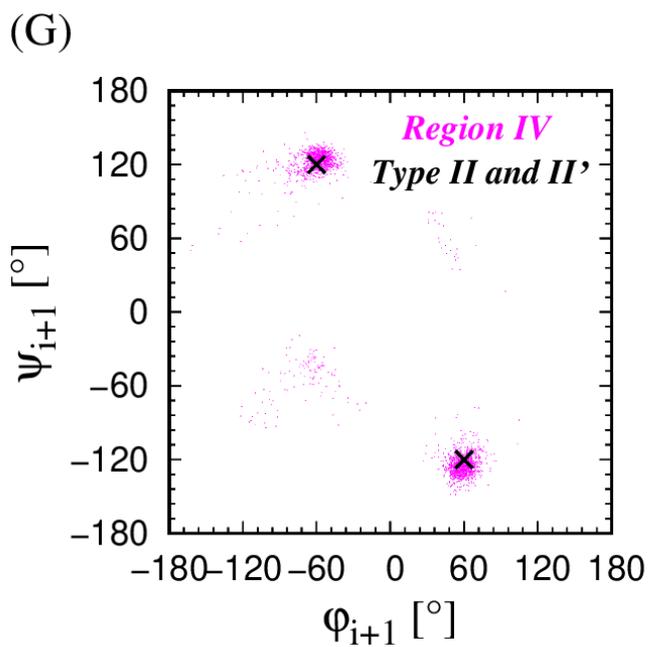

(G)

*Region IV*
*Type II and II'*

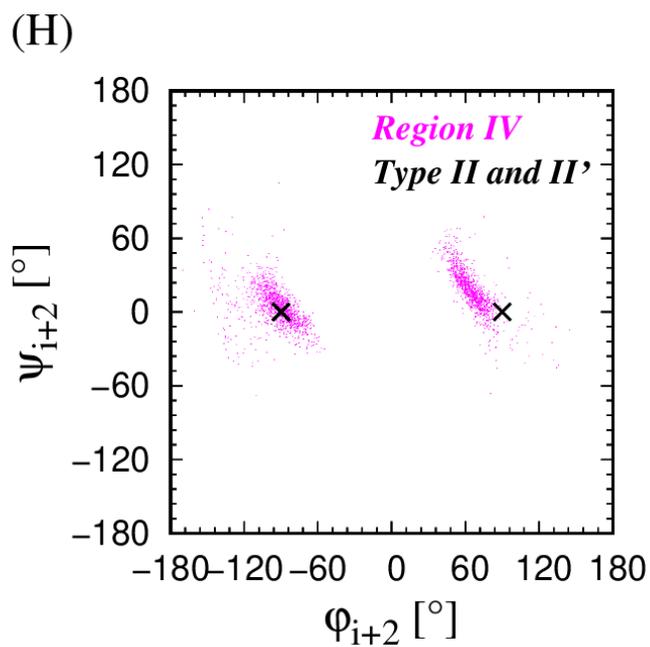

(H)

*Region IV*
*Type II and II'*



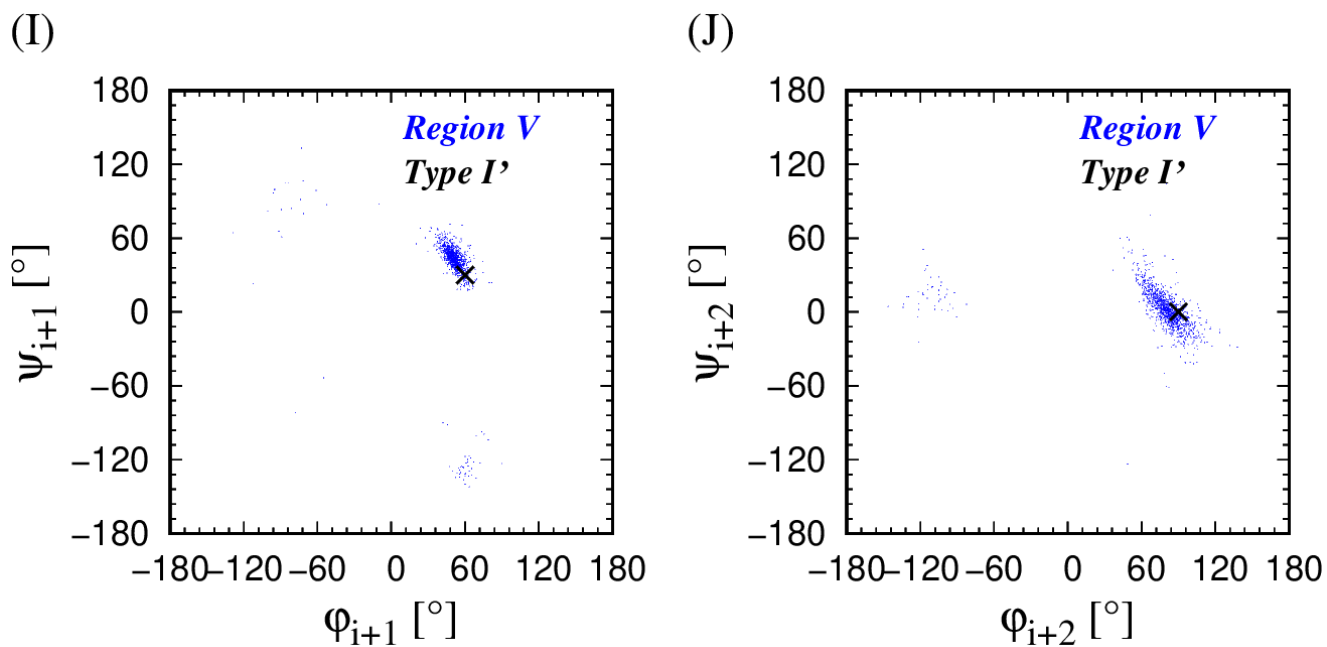

Figure 17: Panels (A-J) show the Ramachandran plots [29] for the beads (i+1) and (i+2) in the middle of the quartet (i,i+1,i+2,i+3) for the five map regions of the (i,i+3) isolated symmetric poking contacts within protein loops. The black X symbols indicate typical average values [3].



|  | Embedded poking contacts | Region I | Region II | Region III | Region IV | Region V |
|---|---|---|---|---|---|---|
| d(i,i+3) [Å] | 5.12 +/- 0.14 | 5.22 ± 0.20 | 5.07 ± 0.29 | 5.36 ± 0.22 | 5.25 ± 0.24 | 5.27 ± 0.17 |
| $\theta_{i+1}$ [°] | 91.35 +/- 2.05 | 90.91± 3.52 | 92.49 ± 4.21 | 108.89±2.75 | 107.39 ± 4.06 | 92.57 ± 3.62 |
| $\theta_{i+2}$ [°] | 91.33 +/- 2.11 | 94.63 ± 3.16 | 101.42±4.81 | 93.60 ± 3.46 | 93.21 ± 4.20 | 94.75 ± 2.95 |
| $\mu$ [°] | 49.78 +/- 3.52 | 48.07 ± 5.62 | 26.22 ± 8.85 | 12.66 ± 7.66 | -12.03 ± 8.34 | -46.96 ± 6.86 |
| $\tau$ [°] | 130.07 +/- 3.37 | 131.42 ± 5.24 | 149.35±6.16 | 153.42±4.70 | 155.00 ± 4.83 | 132.19 ± 6.09 |
| RMSD [Å] | 0.082 +/- 0.039 | 0.135 ± 0.066 | 0.183±0.071 | 0.144±0.074 | 0.172 ± 0.088 | 0.135 ± 0.087 |
| Hydrogen bonded [%] | 99.99 | 97.6 | 86.09 | 96.57 | 96.56 | 98.54 |

Table II: Mean values with standard deviations of various geometrical characteristics calculated for the embedded (i,i+3) symmetric poking contacts, as well as for the five regions of local (i,i+3) isolated symmetric poking contacts located within protein loops. (i,i+3) contact is considered to be hydrogen bonded if at least one hydrogen bond of the type (i,j), (i-1,j),(i+1,j),(i,j-1), or (i,j+1) is present.

## 3.4 Amino acid specificities

We studied the occurrence of 20 amino acid types in the inner positions (i+1) and (i+2) of 176,711 (i,i+3) _embedded symmetric poking contacts_, as well as the most over-expressed (i+1,i+2) amino acid pairs.  We find that there are in total eight amino acids that are over-expressed with the normalized frequencies ranging from 1.12 to 1.56. We find that there is an equal participation of



hydrophilic and hydrophobic amino acids: hydrophilic (GLU, GLN, ARG, LYS) and hydrophobic (ALA, LEU, MET, ILE). More specifically, ALA, LEU, and ILE belong to the aliphatic group (non-polar amino acids with linear or branched sidechains), GLU is negatively charged, ARG and LYS are positively charged, GLN is amidic (polar, uncharged, with R amino group) and MET is a sulfur-containing hydrophobic residue. The same eight amino acids are over-expressed at both positions (i+2) and (i+3). The over-expressed amino acid pairs at positions (i+1,i+2) are made up uniformly from the above over-expressed individual amino acids. We direct an interested reader to Supplementary Information for more details.

Our analysis of the amino acid specificities in the _isolated poking contacts_ in protein loops reveals certain trends: **In Region I (green in the map),** the predominant choice for the (i+1) position is PRO with normalized frequency > 3, whereas for the (i+2) position, the preferred amino acids are ASN and ASP both with normalized frequency of ~3, and the over expressed (i+1,i+2) amino acids pairs are naturally (PRO,ASP) and (PRO,ASN). **Region II (red in the map)** has the same preferences as Region I, with the only difference being a smaller over-expression normalized frequency of the PRO



residue at the position (i+1) dropping to ~2. **In Region III (purple color in the map),** the predominant choice for the (i+1) position is again PRO with an even larger normalized frequency of ~4, whereas GLY is *highly* preferred in position (i+2) with normalized frequency of ~9. The most over-expressed (i+1,i+2) amino acids pairs all have GLY at the position (i+2). **In Region IV (magenta color in the map)**, the predominant choice for the (i+1) position is GLY with normalized frequency of ~4 and PRO follows with a normalized frequency of ~2. In the (i+2) position ASN, GLY, and ASP amino acids are favored with normalized frequencies of ~3. Note the 'reversed' preferences in positions (i+1) and (i+2) for Regions III and IV, that are roughly the geometrical mirror images of each other. **In Region V (blue color in the map),** the mirror image of Region I, the PRO residue is significantly *under-expressed* in the position (i+1) with normalized frequency of only 0.02 [in Region I it was o*ver-expressed* with normalized frequency >3]. The predominant choices for the (i+1) position in **Region V (blue color in the map)** are ASN with normalized frequency of ~5.5, as well as ASP with normalized frequency of ~3.5, while the *highly preferred* amino acid in the position (i+2) is predominantly GLY with normalized frequency of ~10. In all five regions identified



in protein turns, all over-expressed amino acids are hydrophilic. For more detailed information we direct an interested reader to Supplementary Information (Tables SI II to Table SI VII).

We have also studied the amino acid preferences in the 228 local (i,i+3) isolated symmetric poking contacts within protein loops with at least one short bond. We find that in the position (i+1) PRO and ASN residues are preferred having a normalized frequency of ~4, while, for the (i+2) position, we find *overwhelming preference* for a PRO residue with a normalized frequency of ~16. In 75% of all cases, the short bond is the (i+1,i+2) bond.

## 4 Conclusions

While protein secondary structure elements are repetitive and thus iso-directional, the basic role of protein turns is to provide an effective and advantageous way to turn the chain back to itself, changing its direction and thus defining protein boundaries. Turns assist in ensuring the compactness of the globular protein structure. They can be viewed as the structural feature that literally unifies the structure of a globular protein as a whole: helices are made of tight turns locally repeating periodically; non-local β-strands are



coupled by tight turns in hairpin loops; turns delineate the surface of a globu-

lar protein and being predominantly placed at the protein surface necessarily

have a specific amino acid composition; and these turn surface sites are often

implicated in binding and protein function. Protein turns, after the repetitive

protein secondary structure elements, α-helices, and β-sheets, are the most

abundant structural features in proteins.  Here we have demonstrated that

purest protein turns are instances of isolated, local (i,i+3) poking contacts,

that naturally provide a link with our recently presented theoretical frame-

work to understand proteins [30,31]. Our studies presented here allow us to

systematically access, characterize and classify turns in proteins, compre-

hending and unifying numerous approaches and definitions present in the lit-

erature. This unifying perspective of protein turns provides an explanation for

their large structural flexibility and versatility, which allows them to fulfill

successfully all their roles. Even though our analysis is a simple coarse-

grained approach at the level of $C_\alpha$ atoms, it is compatible with the pioneering

studies of turns in the context of Ramachandran maps [29]. Finally, our stud-

ies confirm the notion that turns are emergent snippets of local structures that

are aborted nuclei of helices. They accommodate larger turn angles by pro-



gressively becoming more and more planar. Furthermore, the strict constraint of right-handed helices required by steric clashes for a helix containing many turns is relaxed allowing for turns which can be left-handed and yet carry out the desired goal of turning iso-directional structural entities.

Poking contacts, a measure of the affinity of two parts of a chain, are a vivid example of nature using the same theme in different contexts. Poking contacts with rigid geometrical constraints are local and repetitive in a helix. Distinct geometrical constraints are in play in the formation of sheets assembled from iso-directional strands and, this time, the poking contacts are non-local. The assembly of these building blocks occurs through side-chain mediated interactions, which can be captured in an approximate but harmonious manner by invoking poking contacts between backbone $C_\alpha$ atoms that do not have any geometrical constraints.  There is significant variation in the geometries of the interacting side chains unlike the backbone atoms responsible for the creation of the building blocks. Here we have demonstrated that isolated poking contacts of the backbone atoms play a pivotal role in determining the geometries and types of turns in globular proteins. The recurrent role of poking contacts, a very natural physical manifestation of attraction in the context



of a chain, is an elegant example of nature using every avenue available to her to full effect. Poking contacts are implicated in the formation of helices, sheets, turns, and the assembly of the tertiary structures of globular proteins. From a dynamic point of view, in early folding, there ought to be a profusion of poking (i-i+3) contacts between the backbone atoms, common to all proteins. They ought to be able to form readily under folding conditions in the absence of restrictions on too tight a bending angle or clashes between sidechains because they are entirely local in character. Some of these poking contacts aggregate and become part of a helix, one of the building blocks of proteins. Those that are unable to become part of a coordinated helix can remain as isolated turns fulfilling a useful role in protein structure and function. Thus, protein turns, be they eventually part of a helix or isolated, are heavily involved in the key first steps of the dynamics of folding.

We refer the reader to Supplementary Information for more detailed information.



## Acknowledgments


We are indebted to George Rose for his inspiration and invaluable advice. The computer calculations were performed on the Talapas cluster at the University of Oregon.


## Funding information


This project received funding from the European Union's Horizon 2020 research and innovation program under Marie Skłodowska-Curie Grant Agreement No. 894784 (TŠ). The contents reflect only the authors' view and not the views of the European Commission. JRB was supported by a Knight Chair at the University of Oregon. TXH is supported by The Vietnam Academy of Science and Technology under grant No. NCXS02.05/22-23.


## Conflict of interest

The authors declare that there is no conflict of interest.

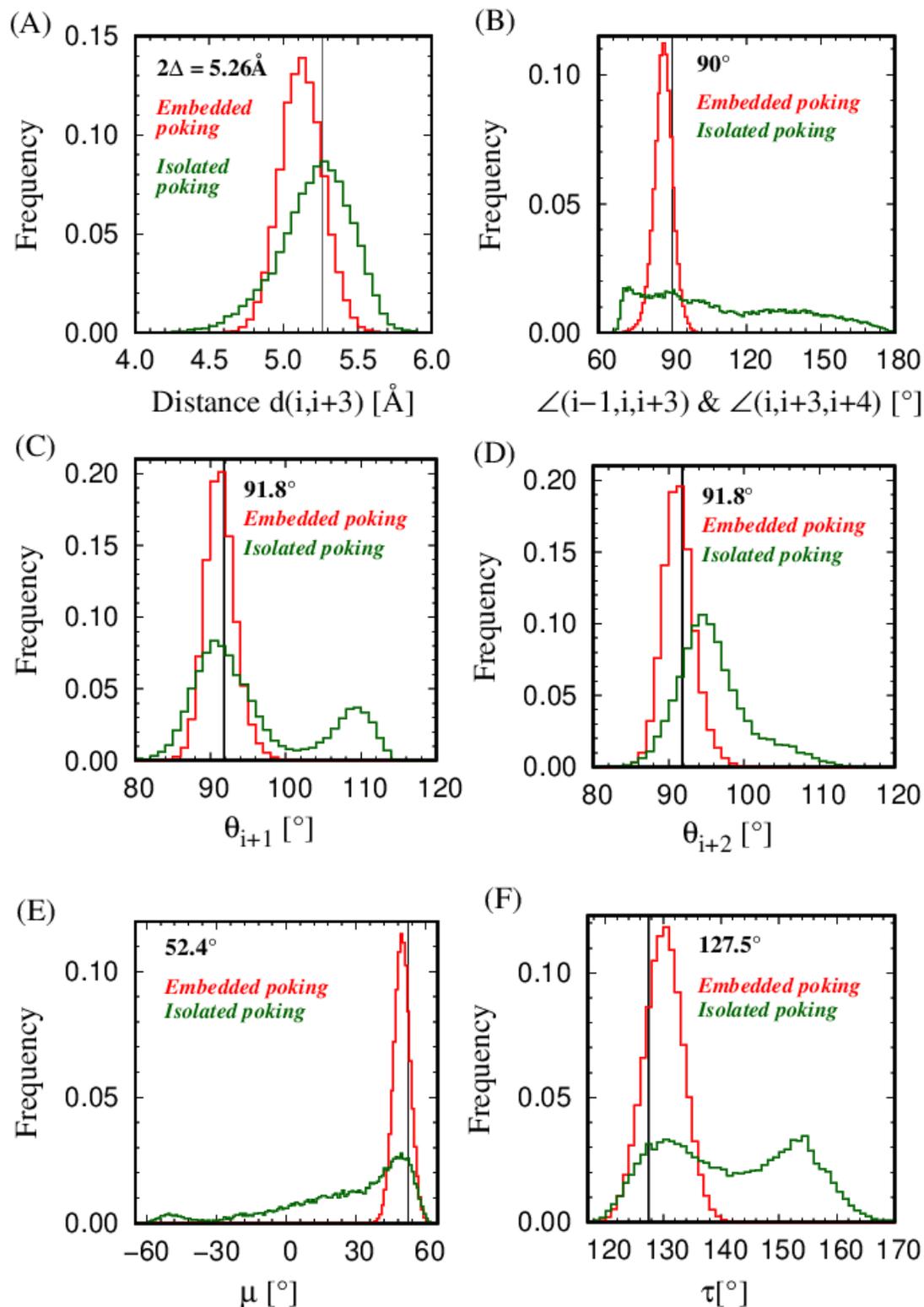



**Figure 1 SI: Geometrical characteristics of embedded and isolated (i,i+3) local poking contacts in our data set comprising 4,391 globular protein chains. There are 176,711 embedded poking contacts for which both the neighboring pairs (i-1,i+2) and (i+1,i+4) are also poking contacts and 21, 571 isolated poking contacts located within protein loops for which neither neighboring pair, (i-1,i+2) or (i+1,i+4), is poking even in an asymmetric manner. An embedded (i,i+3) poking contact is most likely to be within a helix. The red curves depict the histograms for embedded poking contacts, whereas the green curves show the corresponding histograms for isolated poking contacts located within protein loops. The black vertical lines indicate the theoretically predicted values of a given geometrical attribute for the Kepler helix.  Panel (A) shows the distribution of the (i,i+3) distances. The black vertical line indicates the theoretically predicted coin diameter of $2\Delta = 5.26$Å, pertaining to Kepler touching. Panel (B) shows the distribution of the angles (i-1,i,i+3) and (i,i+3,i+4) that are predicted to be 90° (shown as the black vertical line) for embedded poking contacts. Note that although isolated poking contacts in protein loops roughly follow the $2\Delta$ distance constraint of the Kepler helix, they do not necessarily satisfy the 90° angle requirement. Panels (C) and (D) show the distribution of the bond bending angle $\theta_{i+1}$, subtended at point (i+1) by points i and (i+2) and the bond bending angle $\theta_{i+2}$, subtended at point (i+2) by points (i+1) and (i+3 respectively. The black vertical lines in both panels show the ideal value of the bond**



bending angle of 91.8° in the Kepler helix. Panel (E) shows the distribution of the dihedral angle μ for the quartet of beads (i,i+1,i+2,i+3). This is the angle between the planes formed by [i,(i+1), (i+2)] and [(i+1),(i+2),(i+3)]. The vertical black line indicates the value of 52.4° for the Kepler helix. Panel (F) shows the distribution of the turn angle τ for the quartet of beads (i,i+1,i+2,i+3), defined as the angle between the unit vectors along the directions (i,i+1) and (i+2,i+3). The black vertical line shows the value of 127.5°, the turn angle in the Kepler helix.

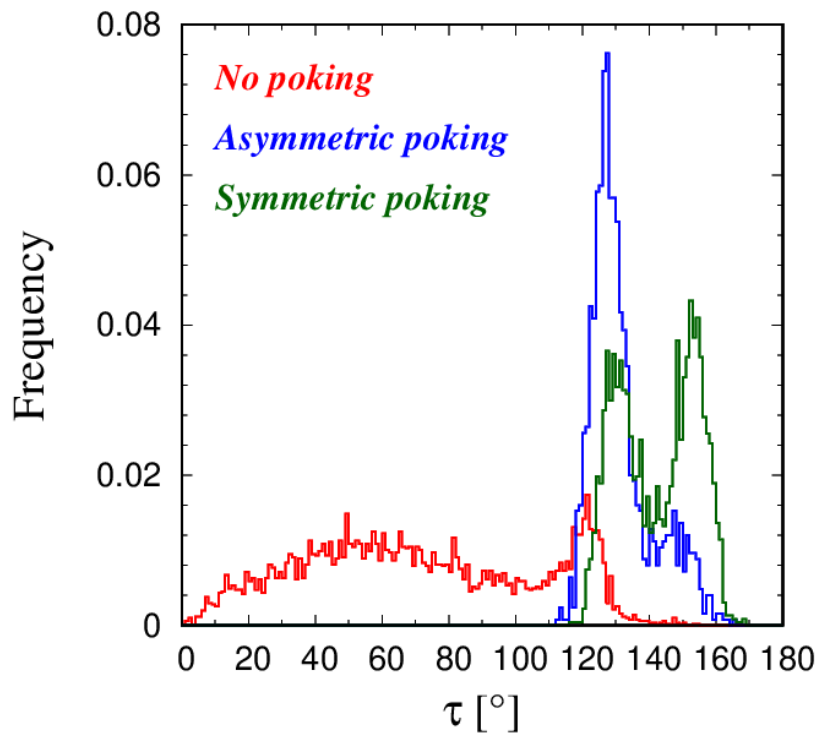

**Figure 2 SI: Frequency distributions of the turn angles τ for three different classes of protein quartets (i,i+1,i+2,i+3). The turn angle τ for a**



quartet is defined as the angle between the unit vectors along the directions (i,i+1) and (i+2,i+3). Here we have considered the quartets coming from the shortest protein loops of length four for which the quartet itself represents the whole loop. There are 8,563 loops of length four in our data set of 4,391 globular protein chains. The histogram in red represents the frequency distribution of the turn angle $\tau$ for 5,048 quartets that do not display poking interaction of any kind between the beads i and i+3. The histogram in blue shows the frequency distribution of turn angles $\tau$ for 1,247 quartets that display asymmetric or 'one-way' poking contact between beads i and i+3. Finally, the histogram in green is the frequency distribution of the turn angle $\tau$ for 2,268 quartets that have (i,i+3) symmetric poking contacts. Note that quartets with symmetric poking contacts between beads i and i+3 are most effective in changing the chain direction.



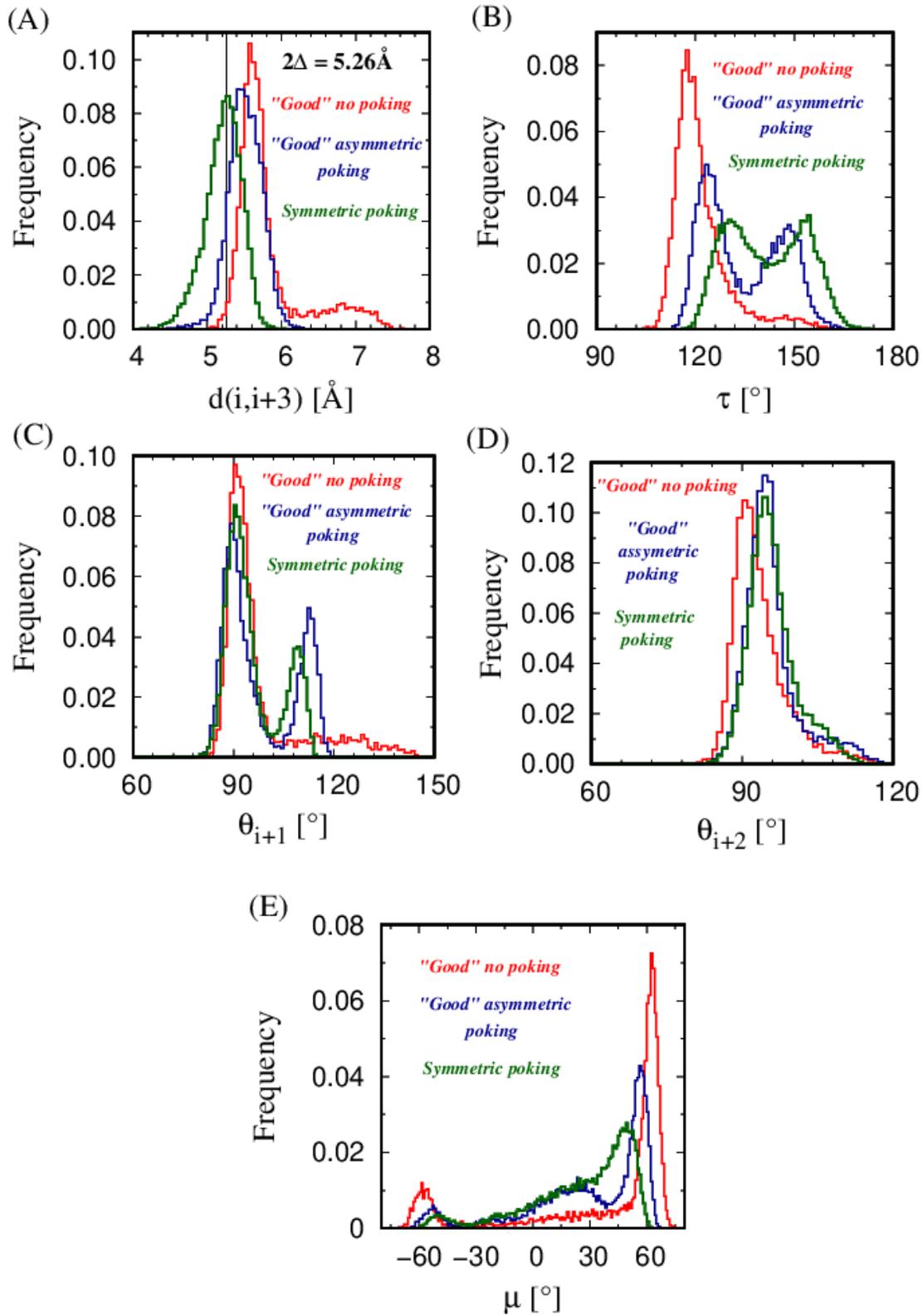



**Figure 3 SI: Geometrical characteristics of three types of 'good' isolated local (i,i+3) contacts found in the 59,464 loops of 4,391 globular protein chains. 21,571 isolated symmetric ('two-way') poking (i,i+3) contacts for which neither neighboring pair, (i-1,i+2) or (i+1,i+4), is poking even in an asymmetric manner (green color); 12,297 asymmetric ('one-way') isolated (i,i+3) poking contacts whose 'effective' distance from the symmetric condition is smaller than 0.263Å (5% of 2Δ) (blue color); and finally 6,500 (i,i+3) pairs without any poking contacts between any pairs of beads (i-1,i+2), (i,i+3), or (i+1,i+4) even within a grace distance 0.263Å (5% of 2Δ) (red color). Panel (A) shows the distribution of the (i,i+3) distances. The black vertical line indicates the theoretically predicted coin diameter of 2Δ = 5.26Å, pertaining to Kepler touching. Panel (B) shows the distribution of the turn angle τ. Panel (C) shows the distribution of the bond bending angle $\theta_{i+1}$, the angle subtended at point (i+1) by points i and (i+2). Panel (D) shows the distribution of the bond bending angle $\theta_{i+2}$, the angle subtended at point (i+2) by points (i+1) and (i+3). Panel (E) shows the distribution of the dihedral angle μ, the angle between the planes formed by [i,(i+1),(i+2)] and [(i+1),(i+2),(i+3)], respectively. The geometrical characteristics of the class of 'good' but not fully developed symmetric poking (i,i+3) contacts (those without any poking and those with asymmetric or 'one-way' poking) gradually tend to the geometrical characteristics of the local (i,i+3) contacts in which**



**symmetric poking is fully established (red and blue histograms tend to the green histograms).**

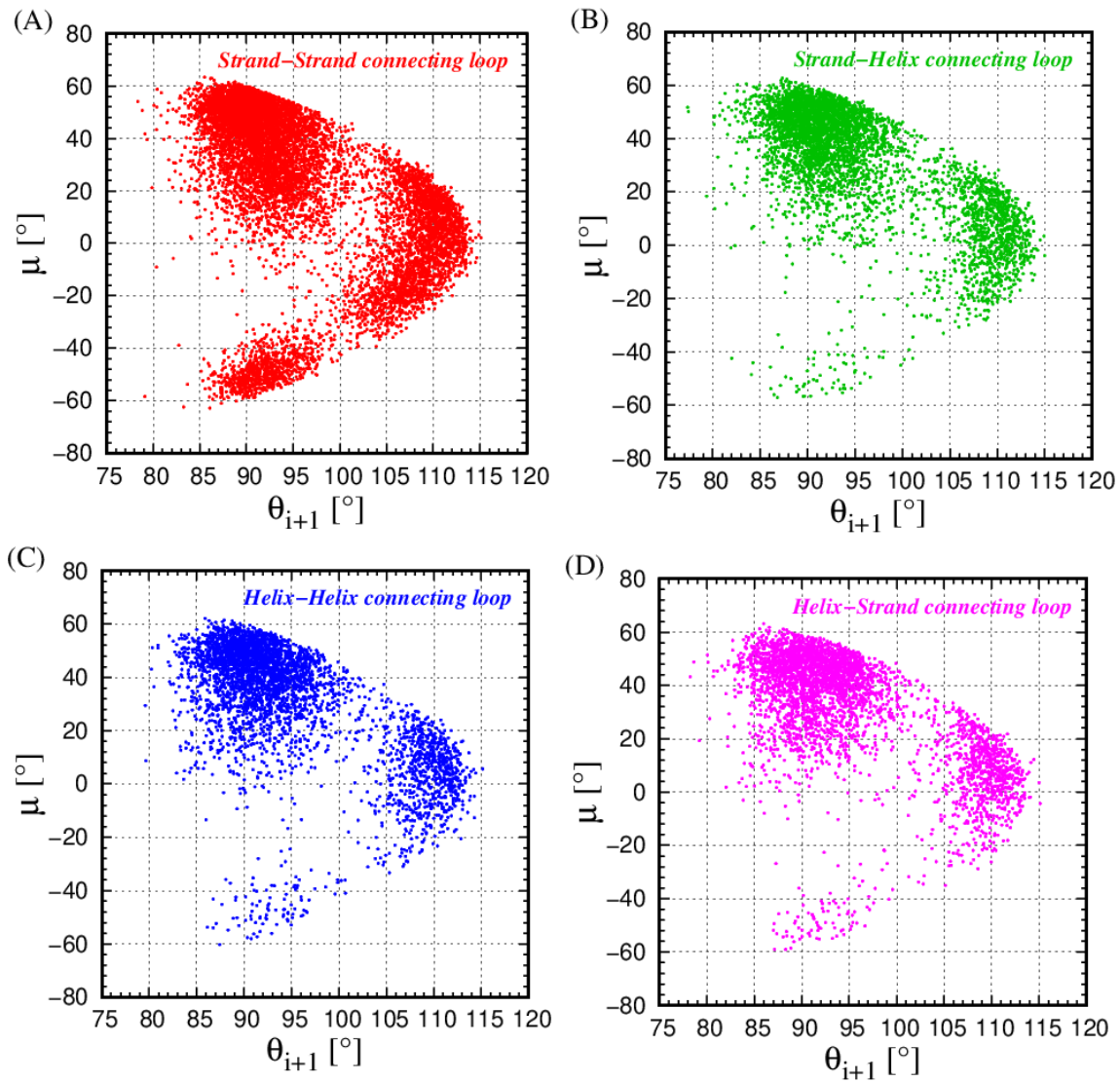

**Figure 4 SI: ($\theta_{i+1}$,$\mu$) cross plots for isolated (i,i+3) symmetric poking contacts located in the loops that connect: a β-strand with a β-strand (9,514 red points) (Panel A); a β-strand with an α-helix (4,493 green points) (Panel B); an α-helix with an α-helix (3,851 blue points) (Panel C); and an α-helix with a β-strand (3,713 magenta points) (Panel D). The**



**cross plots are qualitatively similar underscoring the ability of the same turn types to serve in distinct contexts.**

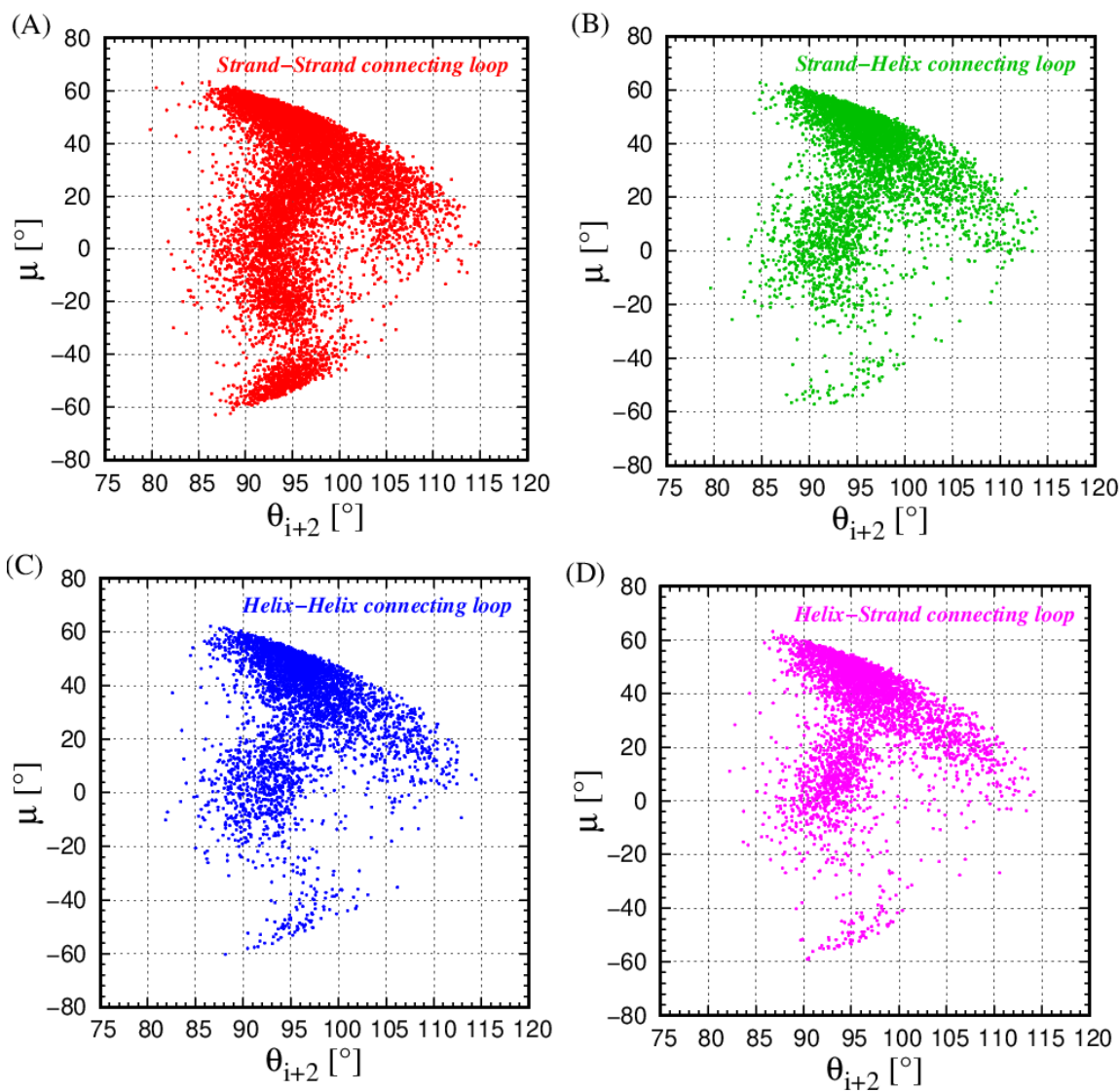

**Figure 5 SI: (θ_{i+2},μ) cross plots for isolated (i,i+3) symmetric poking contacts located in the loops that connect: a β-strand with a β-strand (9,514 red points) (Panel A); a β-strand with an α-helix (4,493 green points) (Panel B); an α-helix with an α-helix (3,851 blue points) (Panel C); and an α-helix with a β-strand (3,713 magenta points) (Panel D). The**



cross plots are qualitatively similar underscoring the ability of the same turn types to serve in distinct contexts.

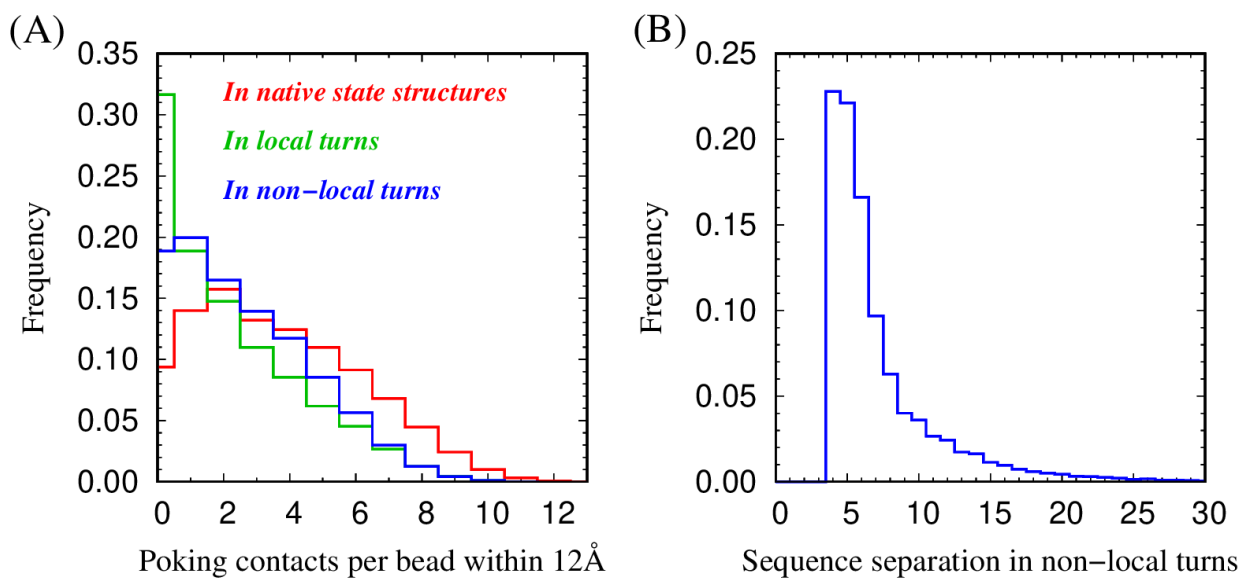

Figure 6 SI: Panel (A) shows the frequency distribution of the total number of symmetric poking contacts (for all sequence separations along the chain) per bead within 12Å in three different cases: for all beads in the native states of 4,391 proteins of our data set (red histogram); for the beads (i+1) and (i+2) of 21, 571 local turns with (i,i+3) isolated symmetric poking contacts (green histogram); and for the beads (i+1) and (i+2) of 23,115 non-local turns identified by means of (i,j) isolated symmetric poking contacts with j>i+3 (blue histogram). We conclude that beads (i+1) and (i+2) are the most 'orphaned' in (i,i+3) isolated protein turns, meaning that they have the smallest number of poking contacts with other parts of the chain, when compared with the non-local (i,j) non-local



**turns, j>i+3. Both local and non-local turns have significantly less poking contacts with other parts of the chain than a typical amino acid anywhere in the protein chain (red histogram lies below the green and blue histograms). Panel (B) shows the frequency distribution of the (j-i) sequence separation of the non-local isolated symmetric poking contact (i,j) within a distance of 6Å. Sequence separations of 4 and 5 comprise approximately 50% of all non-local isolated symmetric poking contacts within a distance of 6Å.**

| Aminoacid type | Aminoacid number | Aminoacid frequency |
|---|---|---|
| LEU | 85,661 | 0.0882 |
| ALA | 82,830 | 0.0853 |
| GLY | 76,106 | 0.0784 |
| VAL | 70,032 | 0.0721 |
| GLU | 63,244 | 0.0651 |
| ASP | 57,797 | 0.0595 |
| SER | 56,931 | 0.0586 |
| LYS | 55,756 | 0.0574 |
| ILE | 54,496 | 0.0561 |
| THR | 53,994 | 0.0556 |
| ARG | 47,203 | 0.0486 |
| PRO | 44,768 | 0.0461 |
| ASN | 42,563 | 0.0438 |
| PHE | 39,166 | 0.0403 |
| GLN | 35,078 | 0.0361 |
| TYR | 35,022 | 0.0361 |
| HIS | 22,608 | 0.0233 |
| MET | 19,714 | 0.0203 |
| TRP | 14,704 | 0.0151 |
| CYS | 13,222 | .0136 |



**Table I SI: Total number and frequencies of occurrences of 20 amino acid types in 970,896 residues in our data set of 4,391 globular protein chains.**

## Embedded (i,i+3) poking contacts

| Aminoacids in (i+1) position | Normalized frequency | Aminoacids in (i+2) position | Normalized frequency | Aminoacid pair (i+1,i+2) | Normalized frequency |
|---|---|---|---|---|---|
| ALA | 1.56 | ALA | 1.62 | ALA - ALA | 2.65 |
| GLU | 1.44 | LEU | 1.46 | GLU - ALA | 2.43 |
| LEU | 1.37 | MET | 1.41 | ALA - LEU | 2.40 |
| GLN | 1.34 | GLN | 1.36 | GLN - ALA | 2.28 |
| MET | 1.31 | ARG | 1.32 | ALA - MET | 2.26 |
| ARG | 1.23 | GLU | 1.31 | GLN - GLN | 2.21 |
| ILE | 1.18 | ILE | 1.25 | LEU - ALA | 2.17 |
| LYS | 1.12 | LYS | 1.19 | MET - ALA | 2.15 |
| VAL | 0.99 | PHE | 0.99 | GLU - LEU | 2.12 |
| PHE | 0.95 | VAL | 0.98 | ALA - ARG | 2.08 |
| TRP | 0.92 | TRP | 0.93 | MET - MET | 2.07 |
| TYR | 0.86 | CYS | 0.91 | LYS - ALA | 2.05 |
| CYS | 0.86 | TYR | 0.88 | ARG - ALA | 2.04 |
| ASP | 0.84 | HIS | 0.79 | GLU - GLU | 2.04 |
| HIS | 0.81 | ASP | 0.74 | LEU - ARG | 2.04 |
| THR | 0.74 | THR | 0.69 | GLN - LEU | 2.03 |
| ASN | 0.68 | ASN | 0.68 | GLU - MET | 2.01 |
| SER | 0.67 | SER | 0.64 | ALA- ILE | 2.00 |
| GLY | 0.43 | GLY | 0.36 | LEU - GLU | 1.99 |
| PRO | 0.17 | PRO | 0.02 | ALA - GLN | 1.98 |

**Table II SI: Normalized frequencies of occurrence of 20 amino acid types in the positions (i+1) and (i+2) of 176,711 (i,i+3) embedded symmetric poking contacts, as well as the top 20 over-expressed (i+1,i+2) amino acid pairs.**



## Isolated (i,i+3) poking contacts: Region I

| Aminoacids in (i+1) position | Normalized frequency | Aminoacids in (i+2) position | Normalized frequency | Aminoacid pair (i+1,i+2) | Normalized frequency |
|---|---|---|---|---|---|
| PRO | 3.37 | ASP | 2.86 | PRO - ASP | 10.86 |
| SER | 1.63 | ASN | 2.45 | PRO - ASN | 8.83 |
| GLU | 1.48 | HIS | 1.51 | LYS - ASP | 6.23 |
| LYS | 1.37 | TRP | 1.38 | PRO - HIS | 4.94 |
| ALA | 1.30 | TYR | 1.32 | PRO - GLU | 4.87 |
| ASP | 1.15 | SER | 1.24 | PRO - TRP | 4.80 |
| ARG | 1.00 | GLN | 1.19 | ALA - ASP | 4.38 |
| GLN | 0.97 | LYS | 1.16 | PRO - SER | 4.30 |
| TRP | 0.89 | PHE | 1.13 | GLU - ASP | 4.24 |
| CYS | 0.80 | GLU | 1.12 | PRO - THR | 4.08 |
| HIS | 0.71 | THR | 0.92 | LYS - ASN | 4.02 |
| THR | 0.70 | ARG | 0.84 | PRO - GLN | 3.95 |
| ASN | 0.68 | CYS | 0.77 | SER - ASP | 3.75 |
| TYR | 0.62 | LEU | 0.68 | GLU - ASN | 3.70 |
| LEU | 0.60 | MET | 0.66 | GLN- HIS | 3.46 |
| MET | 0.59 | ALA | 0.62 | PRO - TYR | 3.40 |
| PHE | 0.58 | GLY | 0.52 | ASP - ASP | 3.37 |
| ILE | 0.48 | ILE | 0.30 | ALA - ASN | 3.32 |
| GLY | 0.45 | VAL | 0.29 | SER - ASN | 3.31 |
| VAL | 0.42 | PRO | 0.22 | TRP - TRP | 3.21 |

**Table III SI: Normalized frequencies of occurrence of 20 amino acid types in the positions (i+1) and (i+2) of 9,604 local (i,i+3) isolated symmetric poking contacts belonging to Region I, as well as the top 20 over-expressed (i+1,i+2) amino acid pairs.**



**Isolated (i,i+3) poking contacts: Region II**

| Aminoacids in (i+1) position | Normalized frequency | Aminoacids in (i+2) position | Normalized frequency | Aminoacid pair (i+1,i+2) | Normalized frequency |
|---|---|---|---|---|---|
| PRO | 1.89 | ASN | 2.46 | PRO - ASN | 5.44 |
| GLU | 1.69 | ASP | 1.99 | PRO - ASP | 5.33 |
| LYS | 1.62 | THR | 1.74 | LYS - TYR | 4.95 |
| ASP | 1.41 | HIS | 1.68 | LYS - ASP | 4.55 |
| SER | 1.40 | TYR | 1.62 | GLU - ASN | 4.52 |
| ALA | 1.29 | PHE | 1.48 | LYS - ASN | 4.36 |
| GLN | 1.16 | TRP | 1.13 | GLU - HIS | 3.99 |
| ARG | 1.02 | SER | 1.11 | SER - CYS | 3.55 |
| THR | 0.92 | CYS | 1.05 | PRO - THR | 3.48 |
| HIS | 0.86 | ARG | 1.03 | ALA - ASN | 3.45 |
| ASN | 0.80 | LYS | 0.94 | SER - ASN | 3.38 |
| TRP | 0.73 | GLU | 0.81 | ARG - TYR | 3.37 |
| TYR | 0.68 | GLN | 0.80 | GLN - TYR | 3.36 |
| VAL | 0.61 | ILE | 0.78 | GLU - THR | 3.36 |
| LEU | 0.60 | VAL | 0.69 | PRO - HIS | 3.35 |
| MET | 0.60 | GLY | 0.59 | TRP - HIS | 3.23 |
| CYS | 0.59 | MET | 0.56 | ARG - HIS | 3.01 |
| ILE | 0.48 | LEU | 0.54 | HIS - ASP | 3.00 |
| GLY | 0.47 | ALA | 0.40 | LYS - THR | 2.97 |
| PHE | 0.45 | PRO | 0.09 | GLU - TYR | 2.92 |

**Table IV SI: Normalized frequencies of occurrence of 20 amino acid types in the positions (i+1) and (i+2) of 5,271 local (i,i+3) isolated**



**symmetric poking contacts belonging to Region II, as well as the top 20 over-expressed (i+1,i+2) amino acid pairs.**

## Isolated (i,i+3) poking contacts: Region III

| Aminoacids in (i+1) position | Normalized frequency | Aminoacids in (i+2) position | Normalized frequency | Aminoacid pair (i+1,i+2) | Normalized frequency |
|---|---|---|---|---|---|
| PRO | 3.67 | GLY | 9.02 | PRO - GLY | 34.25 |
| LYS | 1.79 | ASN | 1.50 | GLU - GLY | 17.06 |
| GLU | 1.63 | ASP | 0.79 | LYS - GLY | 16.85 |
| ASP | 1.04 | CYS | 0.57 | ALA - GLY | 9.87 |
| ARG | 1.03 | HIS | 0.52 | ARG - GLY | 9.81 |
| ALA | 1.02 | TYR | 0.41 | ASP - GLY | 9.09 |
| GLN | 0.95 | PHE | 0.38 | GLN - GLY | 8.94 |
| SER | 0.91 | ARG | 0.37 | VAL - GLY | 8.88 |
| HIS | 0.86 | GLN | 0.34 | SER - GLY | 8.28 |
| VAL | 0.85 | SER | 0.28 | HIS - GLY | 7.59 |
| ASN | 0.70 | LYS | 0.27 | ILE - GLY | 7.26 |
| ILE | 0.69 | GLU | 0.23 | PRO - ASN | 5.81 |
| TYR | 0.65 | THR | 0.21 | ASN - GLY | 5.76 |
| THR | 0.58 | ALA | 0.20 | MET - CYS | 5.42 |
| MET | 0.57 | TRP | 0.18 | THR - GLY | 5.18 |
| TRP | 0.55 | MET | 0.16 | TYR - GLY | 5.15 |
| GLY | 0.54 | LEU | 0.11 | PHE - GLY | 3.92 |
| PHE | 0.49 | VAL | 0.05 | LEU - GLY | 3.71 |
| LEU | 0.47 | PRO | 0.04 | LYS - ASN | 3.68 |
| CYS | 0.18 | ILE | 0.03 | TRP - GLY | 3.40 |

**Table V SI: Normalized frequencies of occurrence of 20 amino acid types in the positions (i+1) and (i+2) of 3,237 local (i,i+3) isolated symmetric**



**poking contacts belonging to Region III, as well as the top 20 over-expressed (i+1,i+2) amino acid pairs.**

## Isolated (i,i+3) poking contacts: Region IV

| Aminoacids in (i+1) position | Normalized frequency | Aminoacids in (i+2) position | Normalized frequency | Aminoacid pair (i+1,i+2) | Normalized frequency |
|---|---|---|---|---|---|
| GLY | 4.31 | ASN | 3.46 | GLY - ASP | 15.79 |
| PRO | 1.64 | GLY | 2.85 | GLY - ASN | 12.03 |
| LYS | 1.13 | ASP | 2.74 | PRO - GLY | 9.02 |
| ASP | 1.00 | CYS | 1.28 | GLY - SER | 8.66 |
| HIS | 0.95 | SER | 1.26 | GLY - THR | 6.30 |
| TYR | 0.91 | THR | 0.82 | LYS - ASP | 5.60 |
| ASN | 0.90 | GLU | 0.80 | HIS - CYS | 5.48 |
| GLU | 0.86 | LYS | 0.77 | LYS - ASN | 5.18 |
| TRP | 0.84 | HIS | 0.73 | GLY - GLU | 4.94 |
| GLN | 0.78 | TRP | 0.72 | LYS - GLY | 4.73 |
| PHE | 0.71 | ARG | 0.68 | GLY - LYS | 4.72 |
| ARG | 0.68 | TYR | 0.54 | TYR - CYS | 4.45 |
| MET | 0.62 | GLN | 0.53 | ASN - ASN | 4.31 |
| ALA | 0.57 | ALA | 0.50 | PRO - ASN | 4.30 |
| SER | 0.49 | PHE | 0.45 | GLY - PRO | 4.21 |
| THR | 0.48 | MET | 0.42 | GLU - ASN | 4.12 |
| LEU | 0.46 | PRO | 0.33 | GLY - ARG | 4.11 |
| CYS | 0.42 | LEU | 0.30 | ASP - GLY | 4.10 |
| VAL | 0.40 | VAL | 0.15 | ASP - ASN | 4.01 |
| ILE | 0.29 | ILE | 0.12 | TYR - ASN | 3.87 |

**Table VI SI: Normalized frequencies of occurrence of 20 amino acid types in the positions (i+1) and (i+2) of 2,298 local (i,i+3) isolated**



**symmetric poking contacts belonging to Region IV, as well as the top 20 over-expressed (i+1,i+2) amino acid pairs.**

## Isolated (i,i+3) poking contacts: Region V

| Aminoacids in (i+1) position | Normalized frequency | Aminoacids in (i+2) position | Normalized frequency | Aminoacid pair (i+1,i+2) | Normalized frequency |
|---|---|---|---|---|---|
| ASN | 5.45 | GLY | 10.24 | ASN - GLY | 57.42 |
| ASP | 3.49 | ASN | 1.32 | ASP - GLY | 34.40 |
| GLY | 1.78 | ASP | 0.57 | GLY - GLY | 17.94 |
| HIS | 1.40 | GLN | 0.29 | LYS - GLY | 13.94 |
| LYS | 1.33 | TRP | 0.28 | HIS - GLY | 13.17 |
| GLN | 1.02 | SER | 0.27 | ARG - GLY | 10.63 |
| GLU | 1.00 | TYR | 0.22 | GLU - GLY | 9.95 |
| ARG | 0.98 | LYS | 0.20 | GLN - GLY | 9.42 |
| ALA | 0.64 | PHE | 0.19 | ASN - ASN | 7.18 |
| SER | 0.54 | CYS | 0.19 | ALA - GLY | 6.56 |
| MET | 0.41 | GLU | 0.18 | SER - GLY | 6.15 |
| TYR | 0.31 | HIS | 0.18 | ASP - ASN | 4.96 |
| PHE | 0.30 | ARG | 0.17 | MET - GLY | 4.20 |
| TRP | 0.23 | THR | 0.12 | HIS - TYR | 4.11 |
| LEU | 0.21 | ALA | 0.06 | ASN - TRP | 3.91 |
| CYS | 0.19 | LEU | 0.05 | TYR - GLY | 3.67 |
| VAL | 0.10 | VAL | 0.04 | GLN - CYS | 3.50 |
| THR | 0.08 | ILE | 0.03 | PHE - GLY | 3.28 |
| PRO | 0.02 | PRO | 0 | ASN - GLN | 3.26 |
| ILE | 0.01 | MET | 0 | GLU - ASN | 2.42 |



**Table VII SI: Normalized frequencies of occurrence of 20 amino acid types in the positions (i+1) and (i+2) of 1,161 local (i,i+3) isolated symmetric poking contacts belonging to Region V, as well as the top 20 over-expressed (i+1,i+2) amino acid pairs.**

### Isolated (i,i+3) poking contacts with 'short' bonds

| Aminoacids in (i+1) position | Normalized frequency | Aminoacids in (i+2) position | Normalized frequency | Aminoacid pair (i+1,i+2) | Normalized frequency |
|---|---|---|---|---|---|
| PRO | 4.089 | PRO | 16.16 | ASN - PRO | 84.63 |
| ASN | 4.00 | ARG | 0.54 | ASP - PRO | 36.81 |
| ASP | 1.77 | TYR | 0.49 | CYS - PRO | 27.95 |
| SER | 1.41 | GLY | 0.45 | HIS - PRO | 24.43 |
| HIS | 1.31 | PHE | 0.44 | MET - PRO | 22.71 |
| CYS | 1.29 | GLU | 0.40 | GLN - PRO | 21.05 |
| MET | 1.26 | HIS | 0.38 | SER - PRO | 20.96 |
| GLN | 1.09 | SER | 0.37 | GLU - PRO | 18.98 |
| VAL | 0.92 | VAL | 0.31 | VAL - PRO | 18.52 |
| GLU | 0.88 | ASN | 0.30 | LYS - PRO | 14.87 |
| ARG | 0.72 | ASP | 0.30 | THR - PRO | 13.70 |
| LYS | 0.69 | TRP | 0.29 | ARG - PRO | 13.70 |
| THR | 0.63 | GLN | 0.24 | TRP - PRO | 12.61 |
| TRP | 0.58 | MET | 0.21 | PRO - PRO | 12.37 |
| ILE | 0.47 | ALA | 0.20 | PRO - ARG | 11.74 |
| ALA | 0.31 | THR | 0.16 | PRO - TYR | 10.58 |
| LEU | 0.20 | LEU | 0.05 | PRO - GLU | 8.76 |
| GLY | 0.17 | LYS | 0 | ILE - PRO | 8.50 |
| TYR | 0.12 | ILE | 0 | PRO - PHE | 7.09 |
| PHE | 0.11 | CYS | 0 | PRO - TRP | 6.30 |



**Table VIII SI: Normalized frequencies of occurrence of 20 amino acid types in the positions (i+1) and (i+2) of 228 protein quartets (i,i+1,i+2,i+3) located in protein loops that have at least one short bond with an isolated symmetric poking contact established between beads i and i+3, along with the top 20 over-expressed (i+1,i+2) amino acid pairs. In 75% of all cases the short bond in question is (i+1,i+2).**